\documentclass[10pt,aps,prd,reprint,superscriptaddress,preprintnumbers,longbibliography,floatfix]{revtex4-2}

\usepackage[utf8]{inputenc}
\usepackage[T1]{fontenc}
\usepackage{anyfontsize}
\usepackage{microtype}

\usepackage{amsmath,amssymb,amsfonts,amsthm}
\usepackage{mathtools}
\usepackage{mathrsfs}
\usepackage{bm}
\usepackage{commath}
\usepackage{tensor}
\usepackage{leftindex}
\usepackage{braket}
\usepackage{accents}
\usepackage{cancel}
\usepackage{siunitx}
\usepackage[normalem]{ulem}
\usepackage{slashed}
\usepackage{graphicx}
\usepackage[dvipsnames]{xcolor}

\usepackage{array}
\usepackage{booktabs}
\usepackage{hhline}
\usepackage{tabularray}

\usepackage{enumitem}
\usepackage{csquotes}
\usepackage{chngcntr}
\usepackage{pifont}

\usepackage{comment}
\usepackage{verbatim}
\usepackage{lipsum} 

\usepackage{xargs}
\usepackage[colorinlistoftodos,textsize=tiny]{todonotes}
\AtBeginDocument{\setlength{\marginparwidth}{\dimexpr(\paperwidth-\textwidth)/2-\marginparsep-4pt\relax}}
\newcommandx{\ssnote}[2][1=]{\todo[linecolor=red,backgroundcolor=red!8,bordercolor=red,textcolor=red,#1]{\textbf{SS:} #2}}

\definecolor{darkcandyapplered}{rgb}{0.64, 0.0, 0.0}
\definecolor{azure(colorwheel)}{rgb}{0.0, 0.5, 1.0}
\definecolor{darkcerulean}{rgb}{0.03, 0.27, 0.49}
\definecolor{darkcyan}{rgb}{0.0, 0.55, 0.55}
\definecolor{darkolivegreen}{rgb}{0.33, 0.42, 0.18}
\definecolor{armygreen}{rgb}{0.29, 0.33, 0.13}
\definecolor{myrtle}{rgb}{0.13, 0.26, 0.12}
\usepackage[
	colorlinks=true,
	linkcolor=darkcerulean,
	citecolor=darkcandyapplered,
	urlcolor=myrtle,
	filecolor=blue,
	bookmarksopen=true
]{hyperref}

\usepackage{placeins}%

\usepackage{orcidlink}
\usepackage{xurl}
\usepackage{longtable}

\DeclareMathAlphabet{\mathpzc}{OT1}{pzc}{m}{it}

\makeatletter
\if@twocolumn

\else

\fi
\makeatother

\renewcommand{\a}{\alpha}

\newcommand{\be}{\begin{equation}}
\newcommand{\ee}{\end{equation}}

\newcommand{\ba}{\begin{eqnarray}}
\newcommand{\ea}{\end{eqnarray}}

\graphicspath{%
  {figs/}%
  {plots/}%
}

\usepackage{cleveref}
\crefname{section}{Sec.}{Secs.}
\Crefname{section}{Section}{Sections}
\crefname{subsection}{Sec.}{Secs.}
\Crefname{subsection}{Section}{Sections}
\crefname{equation}{Eq.}{Eqs.}
\Crefname{equation}{Equation}{Equations}
\crefname{figure}{Fig.}{Figs.}
\Crefname{figure}{Figure}{Figures}
\crefname{table}{Table}{Tables}
\Crefname{table}{Table}{Tables}
\crefname{appendix}{Appendix}{Appendices}
\Crefname{appendix}{Appendix}{Appendices}
\crefname{suppsec}{Sec.}{Secs.}
\Crefname{suppsec}{Section}{Sections}
\crefname{suppsubsec}{Sec.}{Secs.}
\Crefname{suppsubsec}{Section}{Sections}
\usepackage{soul}
\allowdisplaybreaks
\definecolor{db}{rgb}{0.0, 0.0, 0.62}
\definecolor{dm}{rgb}{0.7, 0.01, 0.7}
\definecolor{dr}{rgb}{0.55, 0.0, 0.0}

\newcommand{\Lds}{L_{\mathrm{dS}}}

\usepackage{ulem}
\begin{document}

\title{Quasinormal modes of Kerr-Newman-de Sitter black holes: Linear stability and strong cosmic censorship conjecture}

\author{Subhodeep Sarkar\,\textsuperscript{\,$\sharp$}\,\orcidlink{0000-0001-9015-8837}}
\email{subhodeep.sarkar1@gmail.com}
\affiliation{Department of Physics, Indian Institute of Technology Gandhinagar, Palaj, Gandhinagar 382055, Gujarat, India}

\author{Suvikranth Gera\,\textsuperscript{\,$\sharp$}\,\orcidlink{0000-0001-8597-1719}}
\email{suvikranthg@gmail.com}
\affiliation{School of Physical Sciences, Indian Association for the Cultivation of Science,  2A and 2B Raja S.C. Mullick Road, Kolkata 700032, India}

\author{Avijit Chowdhury\,\textsuperscript{\,$\sharp$}\,\orcidlink{0000-0002-7235-5076}}
\email{avijit.chowdhury@iiap.res.in}
\affiliation{Indian Institute of Astrophysics, Block 2, 100 Feet Road, Koramangala, Bengaluru 560034, India}

\author{Chiranjeeb Singha\,\textsuperscript{\,$\sharp$}\,\orcidlink{0000-0003-0441-318X}}
\email{chiranjeeb.singha@gmail.com}
\affiliation{School of Physical Sciences, Indian Association for the Cultivation of Science, 2A and 2B Raja S.C. Mullick Road, Kolkata 700032, India}

\author{Sreelakshmi Mundakkal Dasan\,\textsuperscript{\,$\sharp$}\,\orcidlink{0009-0005-1818-7285}}
\email{spssmd3290@iacs.res.in}
\affiliation{School of Physical Sciences, Indian Association for the Cultivation of Science, 2A and 2B Raja S.C. Mullick Road, Kolkata 700032, India}

\author{Sumanta Chakraborty\,\orcidlink{0000-0003-3343-3227}}
\email{tpsc@iacs.res.in}
\affiliation{School of Physical Sciences, Indian Association for the Cultivation of Science,  2A and 2B Raja S.C. Mullick Road, Kolkata 700032, India}

\date{\today}
\begin{abstract}
The stability of Cauchy horizons hinges on the decay rate of perturbations of a black hole. For Kerr-Newman-de Sitter black holes, this is determined by the quasinormal modes of coupled gravitational and electromagnetic perturbations, which we compute for the first time. Using two independent numerical methods, we establish the linear stability of the Kerr-Newman-de Sitter black hole. Near extremality, with the electric charge dominating over the rotation, the Cauchy horizon becomes stable, leading to the violation of the strong cosmic censorship conjecture.
\end{abstract}

\maketitle

\begingroup
\csname ltx@footnote@pop\endcsname%
\renewcommand\thefootnote{$\sharp$}
\footnotetext[1]{SS, SG, AC, CS, and SMD contributed equally to this work and should be considered joint first authors.}
\endgroup
\section{Introduction}
Black holes (BHs) have been the cradle for the development of new physics for more than a hundred years \cite{Chandrasekhar:1931ih,Datt1999, Oppenheimer:1939ue, 1968AmSci..56....1W, Raychaudhuri:1953yv,FoursBruhat2016ExistenceTF,Choquetbruhat1969GlobalAO,Penrose:1964wq,Hawking:1976ra,Penrose:1968ar,NarlikarHistoryGR2015,Dadhich2015PostIndependence} and continue to provide new insights into various aspects of gravitational physics \cite{Padmanabhan:2010zzb, Will:2018bme, Berti:2025hly, Ashtekar:2025wnu}.
Through extensive research spanning the past several decades, we have advanced our understanding of black holes, which can be summarised as follows: (a) BHs in general relativity (GR) are very simple --- characterized only by their mass $M$, electric charge $Q$, angular momentum $J$ and cosmological constant $\Lambda$ --- colloquially referred to as the no-hair theorem  \cite{Carter:1971zc, Hawking:1971vc, Israel:1967wq, Robinson:1975bv}; (b) BHs have thermodynamic properties ---  characteristic properties of BHs, such as surface gravity of the event horizon and horizon area can be associated with temperature \cite{Preskill:1992tc, Giddings:1995gd, Chakraborty:2017pmn, Unruh:2017uaw, Mann:2021lif, Almheiri:2020cfm, Schindler:2020pri} and entropy \cite{Bekenstein:1973ur, Bekenstein:1995ju, Sen:1995in, Strominger:1996sh, Rovelli:1996dv, Ashtekar:1997yu, Sen:2012dw}; (c) BHs are linearly stable --- perturbations of the Kerr-de Sitter (KdS) BH (characterized by mass $M$, angular momentum $J$ and cosmological constant $\Lambda$) \cite{Suzuki:1998vy} as well as of the Kerr-Newman (KN) BH (characterized by mass $M$, angular momentum $J$ and electric charge $Q$) \cite{Berti:2005eb, Mark:2014aja, Dias:2015wqa, Dias:2022oqm, Silva:2025khf, Saha:2025nsg, Hintz:2026jiw, Pombo:2026xsj, Destounis:2026ljd} decay exponentially with time and finally (d) asymptotically flat BHs in GR cannot be deformed under static and linear tidal perturbations \cite{Hinderer:2007mb, Damour:2009vw, Binnington:2009bb, Chakraborty:2026qru, Rodriguez:2026iot, Chakraborty:2026kyf} (though the dynamical and non-linear Love numbers are non-zero \cite{Chakraborty:2026qru, Chakraborty:2025wvs,  Bhatt:2024yyz, Saketh:2023bul, Kehagias:2024rtz, Chakraborty:2026dox, Chakraborty:2024gcr, Chakravarti:2025awj}).

Intriguingly, even after such a huge body of literature on various aspects of BH physics, there exist several simple directions that have remained unexplored. We address one such issue in this paper, regarding the linear stability of the Kerr-Newman-de Sitter (KNdS) BHs, under gravito-electromagnetic perturbations, 
for the first time. As a by-product of our stability analysis, we also comment on the stability of the Cauchy horizon, having a direct connection with the deterministic nature of GR.  

The study of linear as well as non-linear stability of a BH has gained much attention lately, since the characteristic modes of oscillation of an isolated BH, akin to the normal modes of a perturbed isolated string, have been observed in most of the binary BH mergers detected by the LIGO-Virgo-KAGRA (LVK) detectors \cite{LIGOScientific:2014wvq, LIGOScientific:2020tif, LIGOScientific:2016lio, LIGOScientific:2025wao}. These characteristic modes, owing to dissipative boundary conditions at the horizon and asymptotically, lead to complex mode frequencies, referred to as the quasinormal modes (QNMs)~\cite{Berti:2025hly, Cardoso:2025npr, Kokkotas:1999bd, Berti:2009kk, Konoplya:2011qq}. The sign of the imaginary part of these QNMs for a BH, under linear/non-linear perturbation, is the governing factor behind the linear/non-linear stability of the BH: if the sign leads to decaying perturbations, the BH is said to be linearly/non-linearly stable.
The determination of these QNMs requires the perturbation equations to separate into radial and angular sectors, which has been achieved for Schwarzschild \cite{Regge:1957td, Vishveshwara:1970zz, Zerilli:1970se}, Reissner-Nordstr\"{o}m \cite{Moncrief:1974ng}, Kerr \cite{Teukolsky:1973ha} and KN \cite{Hintz:2026jiw} BHs. Separable equations for gravitational and electromagnetic perturbations have also been achieved for their associated de Sitter/anti-de Sitter counterparts \cite{Suzuki:1998vy, Dyatlov:2011jd, Zhao:2026gen, Yoshida:2026yly}.

The non-trivial asymptotics induced by a positive and non-zero cosmological constant on BH solutions have interesting consequences, e.g., tidal Love numbers for BHs with de Sitter asymptotics are non-vanishing, while they vanish for the asymptotically flat counterparts \cite{Nair:2024mya}. In the context of stability analysis and QNMs, the implications of a positive and non-zero cosmological constant are much more severe if the BH solution has electric charge. In particular, for the Reissner-Nordstr\"{o}m-de Sitter (RNdS) BH solution, a sufficiently regular extension of perturbations beyond the Cauchy horizon is possible \cite{Brady:1996za, Costa:2014yha, Costa:2017tjc, Cardoso:2017soq, Dafermos:2018tha}. Such a possibility is against the strong cosmic censorship (SCC) conjecture, which effectively asserts that any physically reasonable solutions of Einstein's equations with well-behaved initial data are globally hyperbolic \cite{Simpson:1973ua, Matzner:1979zz, 1982RSPSA.384..301C, Poisson:1990eh, Ori:1991zz, Ori:2000fi, Dafermos:2003wr}. However, the existence of a smoothly penetrable Cauchy horizon, which acts as the boundary of the maximal Cauchy development of initial data, spells doom for the deterministic nature of GR, leading to a violation of the SCC conjecture \cite{Sarkar:2024tkf}. In simple terms, if it is possible to smoothly extend some perturbation modes beyond the Cauchy horizon, the future evolution of those modes cannot be uniquely obtained from the initial data on a Cauchy hypersurface.
After the observation of the violation of SCC conjecture for the RNdS BH in \cite{Cardoso:2017soq}, subsequently it has been extended to several other cases \cite{Konoplya:2008au, Bhattacharjee:2016zof, Dias:2018ufh, Dias:2018etb, Cardoso:2018nvb, Rahman:2018oso, Hod:2018lmi, Hod:2018dpx, Mo:2018nnu, Luna:2019olw, Rahman:2019uwf, Hod:2019zoa, Liu:2019lon, Dias:2019ery, Rahman:2020guv, Casals:2020uxa, Hod:2020ktb, Bhattacharjee:2020gbo,Bhattacharjee:2020nul, Emparan:2020rnp, Singha:2022bvr, Davey:2022vyx, Davey:2024xvd, Deng:2026gtp}. In this paper, we explore the case of KNdS BH spacetime, focusing on the gravito-electromagnetic perturbations.  

\section{Gravito-electromagnetic perturbations of Kerr--Newman--de Sitter spacetime}
In this section, we will discuss the separability of the radial and angular perturbation equations for gravitational and electromagnetic perturbations of a KNdS BH, which in the Boyer--Lindquist coordinates, reads
\begin{align}
ds^{2}={}&\rho^{2}\left(\frac{dr^{2}}{\Delta_{r}}+\frac{d\theta^{2}}{\Delta_{\theta}}\right)-\frac{\Delta_{r}}{(1+\alpha)^{2}\rho^{2}}\left(dt-a\sin^{2}\theta\,d\phi\right)^{2}\nonumber\\
&+\frac{\Delta_{\theta}\sin^{2}\theta}{(1+\alpha)^{2}\rho^{2}}\left[a\,dt-(r^{2}+a^{2})\,d\phi\right]^{2}\,,
\label{eq:metricKNdS}
\end{align}
where, $\rho^{2}\equiv r^{2}+a^{2}\cos^{2}\theta$, $\Delta_{\theta}\equiv 1+\alpha\cos^{2}\theta$, along with,
\begin{equation}
\begin{split}
\Delta_{r}&\equiv\Delta_{0}-\frac{\alpha}{a^{2}}\,r^{2}(r^{2}+a^{2})\,,\\
\Delta_{0}&\equiv r^{2}-2Mr+a^{2}+Q^{2}\,,\qquad \alpha\equiv\frac{\Lambda a^{2}}{3}\,.
\end{split}
\label{eq:metric_variables_1}
\end{equation}
The above metric describes a BH spacetime for a certain region of the four-dimensional parameter space spanned by $(M,Q,a,\Lambda)$~\cite{Stuchlik:2000pla, Franzen:2020gke, Park:2023lfc}; see Appendix~\ref{sec:AppA} for details on this parameter space.

The KNdS spacetime typically has three horizons, determined by the zeros of $\Delta_{r}$: the event horizon $r_{+}$, the Cauchy horizon $r_{-}$ and the cosmological horizon $r_{\rm c}$. In what follows, we will always consider sub-extremal configurations with $r_{-}<r_{+}<r_{\rm c}$, so that all three horizons exist and are distinct. 

By taking different limits of \cref{eq:metricKNdS}, we can obtain several other well-known BH solutions in GR, e.g., setting $Q=0$, we get back the KdS BH, while by setting $\Lambda=0$, we get the KN BH solution. For the KdS BH, the perturbation equations separate neatly into radial and angular parts, for massless fields with integer $(s=0,1,2)$ and half-integer $(s=1/2,3/2)$ spins \cite{Suzuki:1998vy}. In contrast, for gravito-electromagnetic perturbations of a KN BH, such a separability between radial and angular parts has only been discovered recently \cite{Hintz:2026jiw}; earlier studies treated the rotation perturbatively~\cite{Pani:2013ija, Pani:2013wsa}. Following this, the separability of gravito-electromagnetic perturbations into radial and angular parts, for a KNdS BH has also been achieved in \cite{Zhao:2026gen, Yoshida:2026yly} (even though, for the KNdS BH, the separability of the perturbation equations into radial and angular parts was already known for massless fields with $s=0,1/2$ \cite{Suzuki:1998vy}). 

For a KNdS BH, we obtain the following set of two coupled first order differential equations, describing gravito-electromagnetic perturbations, for the radial and angular sectors, respectively, 
\begin{align}
\frac{d \mathcal{R}^{(\varsigma)}_{1,2}}{dr}&=\pm\left(\lambda+\alpha_{r}\right)\frac{\mathcal{R}^{(\varsigma)}_{1,2}}{\sqrt{\Delta_{r}}}+\left(R_{0}^{(\varsigma)}\mp i R_{1}^{(\varsigma)}\right)\frac{\mathcal{R}^{(\varsigma)}_{2,1}}{\sqrt{\Delta_{r}}}\,,
\label{eq:Firstorder_radial}
\\
\frac{dS^{(\varsigma)}_{1,2}}{d\theta}&=\pm i\left(\alpha_\theta-\lambda\right) \frac{S^{(\varsigma)}_{1,2}}{\sqrt{\Delta_\theta}}\mp i\left( \Theta_1^{(\varsigma)}\mp i \Theta_0^{(\varsigma)}\right)\frac{S^{(\varsigma)}_{2,1}}{\sqrt{\Delta_\theta}}\,,
\label{eq:Firstorder_angular}
\end{align}
where the upper signs go with the first index, $\lambda$ is the separation constant and $\varsigma=1$ describes the positive spin system, while $\varsigma=-1$ depicts the negative spin systems. 

\begin{figure*}[htbp]
    \centering
    \includegraphics{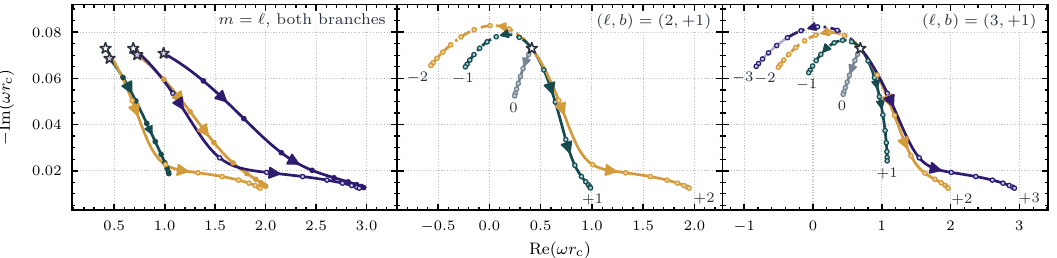}
    \caption{\textbf{Photon sphere modes of a KNdS BH:} The plot depicts the photon sphere modes of a KNdS BH, with $y_{+}=1/2$ and $\mathcal R=0.95$, as $\Theta=\arctan(a/Q)$ increases (direction of arrowheads) from the RNdS modes at $\Theta=0$ (stars, filled for $b=-1$ and hollow for $b=+1$) to generic $\Theta$. \textbf{Left:} the co-rotating mode frequencies are presented with $m=\ell$. \textbf{Other panels:} the frequencies of the multiplets $(\ell,b)$ are depicted.}
    \label{fig:ps_modes_tracks_by_ell}
\end{figure*}

In the above equations, we have introduced two radial functions, $R_{0}^{(\varsigma)}(r)$ and $R_{1}^{(\varsigma)}(r)$, along with two angular functions, $\Theta_{0}^{(\varsigma)}(\theta)$ and $\Theta_{1}^{(\varsigma)}(\theta)$, all of which have the following expressions,
\begin{widetext}
\begin{align}
&R_{0}^{(\varsigma)}\equiv \frac{3\Delta_{r}'}{4\sqrt{\Delta_{r}}}\left(\frac{\nu}{\nu_{0}}\frac{\mathcal{S}_{0}}{\mathcal{S}_{r}}\right)
+\frac{i\varsigma(1+\alpha)\nu_{0}}{\sqrt{\Delta_{r}}\nu}\Bigg[\left\{\frac{\mathcal{S}_{r}}{\mathcal{S}_{0}}(r^{2}+a^{2})\omega-am\frac{\mathcal{S}_{0}}{\mathcal{S}_{r}}\right\}
+\frac{\mathcal{S}_{r}}{\mathcal{S}_{0}}\left(\frac{2Q^{2}\Delta_{0}\omega}{\nu_{0}}\right)\Bigg]+\widetilde{R}_{0}^{(\varsigma)}~,
\\
&R_{1}^{(\varsigma)}\equiv \frac{3iM}{2Q}\,\frac{\mathcal{S}_{0}}{\mathcal{S}_{r}}
-\frac{2\varsigma(1+\alpha)Q(Mr-Q^{2})\,\omega}{\nu}\,\frac{\mathcal{S}_{r}}{\mathcal{S}_{0}}\sqrt{\frac{\Delta_{0}}{\Delta_{r}}}+\widetilde{R}_{1}^{(\varsigma)}~,
\\
&\Theta_0^{(\varsigma)}\equiv \frac{3}{2}\cot\theta\,\frac{\sqrt{H_\theta}}{\sqrt{H_{0}\Delta_\theta}}
+\varsigma(1+\alpha)\sqrt{\frac{H_{0}}{H_{\theta}}}\bigg[\frac{m}{\sin\theta\sqrt{\Delta_{\theta}}}
-\frac{a\omega\left(M^{2}+Q^{2}\right)\sin\theta}{\nu}\bigg]+\widetilde{\Theta}_{0}^{(\varsigma)}~,
\\
&\Theta_1^{(\varsigma)}\equiv-\frac{3M}{2Q}\,\frac{\sqrt{H_\theta}}{\sqrt{H_{0}\Delta_\theta}}
-\frac{2}{\nu}aQ\cos\theta\varsigma(1+\alpha)M\omega\sqrt{\frac{H_{0}}{H_{\theta}}}+\widetilde{\Theta}_{1}^{(\varsigma)}~.
\end{align}
\end{widetext}
The details of the derivation, leading to the first order equations can be found in Appendix~\ref{Separability} and the explicit definitions of various quantities appearing in the above equations are presented in Appendix~\ref{App:exp}.

In general, the analytical determination of the separation constant $\lambda$ is non-trivial, but in the static limit and for a small cosmological constant, it takes the following simple form,
\begin{align}
\lambda=b\sqrt{(\ell-1)(\ell+2)+\left(\frac{3M}{2Q}\right)^{2}}+\mathcal{O}(\alpha)\,,
\end{align}
where $b=+1$ describes the gravitational-led and $b=-1$ the electromagnetic-led perturbations, with $\ell\geq2$. The explicit expression for the $\mathcal{O}(\alpha)$ term can be found in Appendix~\ref{sec:lambda}. Note that our equations above differ from those in \cite{Yoshida:2026yly, Zhao:2026gen}: since the quantities with a tilde vanish as $\Lambda\to0$, they reduce to the corresponding KN equations of \cite{Hintz:2026jiw}, which was not the case for \cite{Yoshida:2026yly, Zhao:2026gen}.

\section{Gravito-electromagnetic quasinormal modes of Kerr--Newman--de Sitter black holes}
We now compute the gravito-electromagnetic QNMs from the separated radial and angular equations, \cref{eq:Firstorder_radial,eq:Firstorder_angular}, for perturbations with time and azimuthal dependence $\exp(-i\omega t+im\phi)$ and spin label $\varsigma=1$. A KNdS BH is characterized by four parameters, $M$, $a$, $\Lambda$ and $Q$. To traverse this parameter space, we follow~\cite{Davey:2024xvd} and use the spherical polar parametrization $\{y_{+},\mathcal{R},\Theta\}$, where,
\begin{equation}
y_{+}\equiv\frac{r_{+}}{r_{\rm c}}\,;
\qquad 
\mathcal{R}^{2}\equiv \frac{r_{-}}{r_{+}}\,;
\qquad 
\tan\Theta \equiv \frac{a}{Q}\,.
\label{eq:sphpolparam}
\end{equation}
Here $y_{+}\in[0,1]$ is the off-Nariai parameter, with the Nariai limit being $y_{+}=1$; $\mathcal R\in[0,1]$ is the off-extremality parameter, with the extremal limit being $\mathcal R=1$; and $\Theta\in[0,\pi/2]$ measures the ratio of rotation to charge~\cite{Davey:2024xvd}. For fixed $y_{+}$ and $\mathcal R$ values, $\Theta$ moves from the RNdS BH at $\Theta=0$ to the KdS BH at $\Theta=\pi/2$, as $M$, $a$, $Q$ and $\Lambda$ all change. 

We solve the separated radial and angular equations by the method of direct integration~\cite{Chandrasekhar:1975zza, Pani:2013pma} (Appendix~\ref{sec:directint}), as well as with a modified spectral method~\cite{Cardoso:2013pza, Dias:2015nua, Jansen:2017oag, Chung:2023wkd} (Appendix~\ref{sec:numericspectral}). The spectral method shows exponential convergence and the two methods agree very closely, which provides a strong consistency check on our numerical results (Appendix~\ref{sm_sec:convergence}); the figures make use of the results obtained from the spectral method. The QNMs of KNdS BHs can be broadly classified into three classes: (a) photon sphere modes, (b) near-extremal (NE) modes, and (c) de Sitter (dS) modes \footnote{Note that, in all the plots below, we have labeled and segregated the QNMs as `photon sphere', `NE' or `dS' modes by tracking them back to the limit $\Theta\to0$ (indicated by the stars in the figures) and then identifying with the RNdS counterpart. Thus, each mode is characterized by the values of $(\ell,b,m,n)$ that its RNdS counterpart stands for.}. Among these, the photon sphere and near-extremal modes are depicted here; a discussion on various features of the QNM spectrum, including the convergence and the dS family of modes, can be found in Appendices~\ref{sec_sm:features_qnm_specturm}, \ref{sm_sec:convergence}, and \ref{app:dS_modes}.

\begin{figure*}[thbp]
    \centering
    \includegraphics[width=0.9\textwidth]{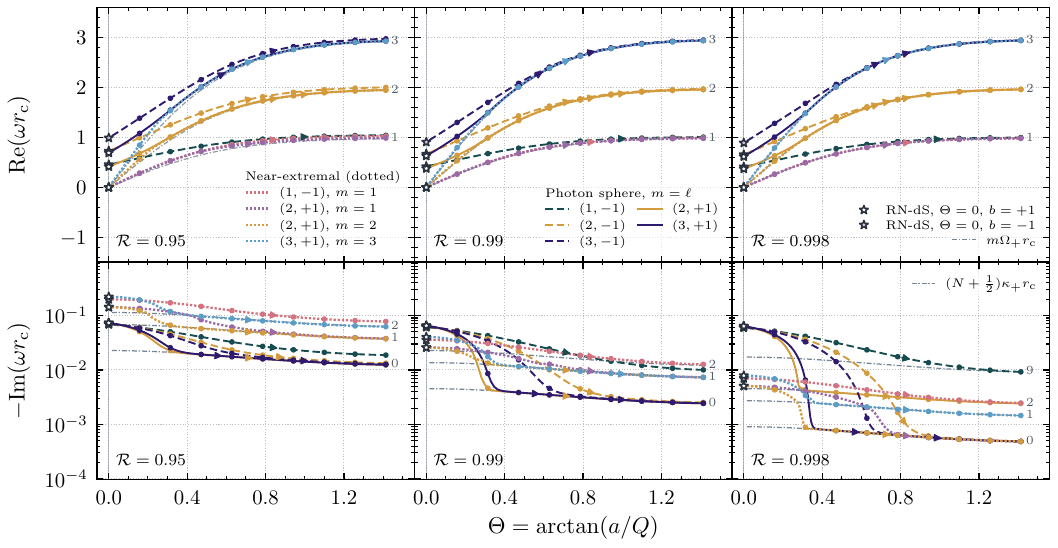}
    \caption{\textbf{Real and imaginary parts of the NE and photon sphere modes} for $y_{+}=1/2$ and $\mathcal R=0.95$, $0.99$ and $0.998$, respectively (left to right). As evident, the $-\mathrm{Im}(\omega r_{\rm c})$ asymptotes to $(N+1/2)\kappa_+r_{\rm c}$ for large $\Theta$. The legends are shared across all six panels and the arrowheads point towards increasing $\Theta$.}
    \label{fig:re-im-three-R}
\end{figure*}

We start with the fundamental QNM frequencies associated with the photon sphere modes, computed using the spectral method in \cref{fig:ps_modes_tracks_by_ell}. The QNMs are computed for various choices of the parameter $\Theta$, at fixed $y_{+}=1/2$ and $\mathcal{R}=0.95$, and are presented as trajectories in the complex $\omega$-plane. In \cref{fig:ps_modes_tracks_by_ell}, besides the photon sphere modes of the KNdS BH, we also mark the QNMs of the RNdS BH~\cite{Stuchlik:2025mjj} associated with the same set of $(y_{+},\mathcal{R})$ values, obtained by taking the $\Theta\to 0$ limit. As is evident, all the QNMs of the KNdS BH associated with different $m$ values branch out from the RNdS QNMs, which is analogous to the Zeeman splitting observed in the hydrogen atom. Moreover, the imaginary parts of the QNM frequencies are negative for all $(\ell,m,b)$ modes, signalling the linear stability of the KNdS BH. Another interesting feature emerging from \cref{fig:ps_modes_tracks_by_ell} corresponds to the existence of a certain $\Theta$ value for which $|\textrm{Im}\,\omega_{\rm QNM}|$ is the largest; this fastest decaying mode is always found in the negative-$m$ branches for both gravitational and electromagnetic perturbations. On the other hand, the positive-$m$ branches show no such behaviour. 

The arrowhead in \cref{fig:ps_modes_tracks_by_ell} depicts the evolution of the fundamental ($n=0$) photon sphere modes in the complex frequency plane as $\Theta$ increases. The spectrum of these modes clearly bifurcates from the RNdS value into two distinct families: the electromagnetic-led ($b=-1$) and gravitational-led ($b=+1$) branches, each of which traces entirely different topological paths. 
Selected values of the QNM frequencies for the photon sphere modes have been tabulated in \cref{tab:gem-complete-spectrum} of Appendix~\ref{app:numerical_values_gem_qnms}. 

\begin{figure*}[thbp]
    \centering
    \includegraphics{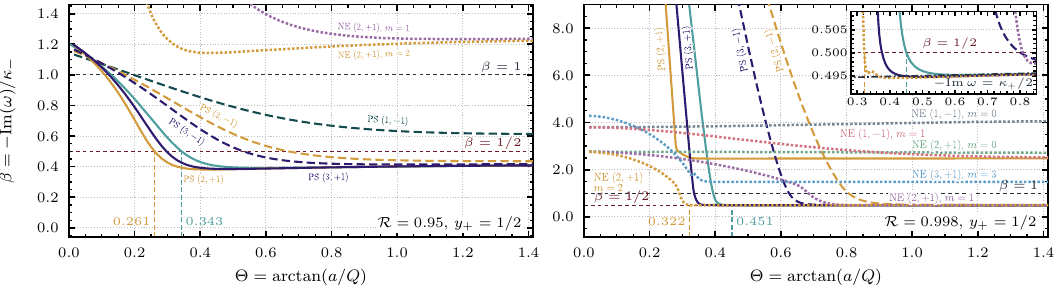}
    \caption{\textbf{Strong cosmic censorship for KNdS BHs} with $y_{+}=1/2$. We show $\beta=-(\mathrm{Im}\,\omega/\kappa_{-})$ for the QNMs. \textbf{Left:} $\mathcal R=0.95$; the vertical lines mark where the photon sphere modes cross the $\beta=1/2$ line, leading to the violation of the SCC conjecture. \textbf{Right:} $\mathcal R=0.998$; the vertical lines mark where the NE $(2,+1)$ mode with $m=2$ and the eikonal curve cross $\beta=1/2$. Here the least damped mode at small $\Theta$ is the NE mode and violates the SCC conjecture.}
    \label{fig:scc-gem}
\end{figure*}

The NE modes, on the other hand, have been presented in \cref{fig:re-im-three-R}. Since the NE modes become important for large $\mathcal{R}$, i.e., close to the extremal configuration, we have shown the variation of the real and imaginary parts of the frequencies of the NE modes for three different choices of $\mathcal{R}=0.95, 0.99$ and $0.998$, respectively. In all these cases $y_{+}$ has been fixed to $1/2$. As evident from \cref{fig:re-im-three-R}, it follows that the real part of the NE mode frequencies is not very sensitive to the choice of $\mathcal{R}$, and asymptotes to those of photon sphere modes as $\Theta$ increases, and they approach $m\Omega_{+}$ in the KdS limit, where $\Omega_+=a/(r_+^2+a^2)$ is the angular velocity of the event horizon. The imaginary part of the NE mode frequencies depicts a very interesting behavior --- as the BH nears more and more towards extremality by increasing $\mathcal{R}$, gradually, the NE modes start to dominate over the photon sphere modes, i.e., for large enough $\mathcal{R}$, or, very close to extremality, the lowest lying QNM is always a NE mode, rather than a photon sphere mode. In particular, there exists some intermediate value of $\mathcal{R}=\mathcal{R}_{\rm c}$, for specific $(\ell,b,m)$ values, at which the imaginary parts of the NE and photon sphere mode frequencies become identical, beyond which, the NE modes decay slower and hence dominate over the photon sphere modes. Note that the imaginary part of the NE mode frequency, unlike the real part, is always different from the photon sphere modes even in the KdS limit. 

Another intriguing feature of the NE modes, clearly seen from \cref{fig:re-im-three-R}, is the result that in the limit of large $\mathcal{R}$ and large $\Theta$, the imaginary part of the QNM frequencies asymptotes to $-(N+1/2)\kappa_{+}$, where $N\in \mathbb{I}^{+}$ and $\kappa_{+}$ is the surface gravity of the event horizon. In fact, \cref{fig:re-im-three-R} also depicts that the imaginary parts of the frequencies of photon sphere modes have the same asymptotic behaviour, but the value of $N$ is different. A similar behaviour has also been observed for scalar field perturbations in KN \cite{Davey:2023fin} and KNdS \cite{Davey:2024xvd} BHs. For example, consider the $(2,+1)$ mode with $m=2$, for which the imaginary part of the lowest lying NE mode asymptotes to $N=1$, while the imaginary part of the photon sphere mode asymptotes to $N=0$, at $\mathcal R=0.95$ and for large $\Theta$. This also holds for $\mathcal{R}=0.99$ as well, even though the imaginary parts of the NE modes take smaller values compared to the $\mathcal R=0.95$ case, for generic $\Theta$. For $\mathcal R=0.998$, on the other hand, the imaginary part of the $(2,+1)$ mode with $m=2$, asymptotes to $N=0$ for the NE sector, while the imaginary part of the photon sphere mode asymptotes to $N=2$, see \cref{fig:re-im-three-R}. Thus, the NE and photon sphere modes switch their roles close to extremality, with the NE mode becoming the least damped one,  which is also the case for scalar perturbations, as found in \cite{Dias:2022oqm, Davey:2024xvd}. 

\section{Implications for the strong cosmic censorship conjecture}
SCC conjecture is intimately connected to the existence of a perturbation that can be extended beyond the Cauchy horizon in a sufficiently regular manner. The possibility of such a regular extension arises from the comparison between the late-time decay and the infinite blueshift experienced by the ingoing perturbation modes near the Cauchy horizon, which is characterized by the spectral gap parameter $\beta=(|\textrm{Im}\omega|/\kappa_{-})$,where $\kappa_{-}$ denotes the surface gravity of the Cauchy horizon. For $\beta<(1/2)$, the perturbations are irregular at the Cauchy horizon, and Christodoulou's formulation of the SCC conjecture is respected, whereas for $\beta>(1/2)$, a sufficiently regular extension beyond the Cauchy horizon is possible, signalling a violation of the SCC conjecture \cite{Hintz:2015jkj, Cardoso:2017soq}. We explore the variation of $\beta$ with the parameter $\Theta=\arctan(a/Q)$ below.

Our results from \cref{fig:scc-gem} demonstrate that, within the parameter space explored in this work, irrespective of the choice of the extremality parameter $\mathcal{R}$, the SCC conjecture is violated for $\Theta<\Theta_{\rm c}$, where the least damped mode (either, photon sphere or, NE) crosses $\beta = 1/2$ line.
For smaller values of $\mathcal{R}$, we note from \cref{fig:scc-gem} that increasing the $\Theta$ parameter drives the photon sphere modes toward the region below the $\beta=(1/2)$ line, thereby favouring the validity of the SCC conjecture in the KdS limit. On the other hand, decreasing $\Theta$, can lead to the photon sphere modes to cross the $\beta=(1/2)$ critical line, resulting in a violation of the SCC conjecture for the RNdS case. This behaviour indicates that a large electric charge, along with a positive cosmological constant, plays an important role in facilitating violations of the SCC conjecture, whereas rotation tends to counteract this effect. In fact, for each choice of $(y_{+},\mathcal{R})$, there exists a critical $\Theta_{\rm c}$, such that for $\Theta>\Theta_{\rm c}$, the SCC conjecture is respected, while for $\Theta<\Theta_{\rm c}$, it is violated. For example, the choice of $(y_{+}=0.5,\mathcal{R}=0.95)$, yields $\Theta_{\rm c}=0.261$, while for the choice $(y_{+}=0.5,\mathcal{R}=0.998)$, we obtain $\Theta_{\rm c}=0.322$. In summary, there will always be a region $0\leq \Theta<\Theta_{\rm c}$, where the SCC conjecture will be violated for the KNdS BH. 

Similarly, there exists a critical value of the extremality parameter $\mathcal{R}_{\rm c}\sim 0.995$, such that for $\mathcal{R}<\mathcal{R}_{\rm c}$, the photon sphere modes are the least damped, while for $\mathcal{R}>\mathcal{R}_{\rm c}$, the NE modes start to dominate the QNM spectrum, with at $\mathcal{R}=\mathcal{R}_{\rm c}$, the imaginary parts of the $(2,+1)$ photon sphere and NE modes cross each other. 
This behavior can be quantitatively understood as follows: for $\mathcal{R}<\mathcal{R}_{\rm c}$, as $\Theta$ increases, the QNM frequencies of the co-rotating photon sphere modes approach the following form $\omega\simeq m\Omega_{+}-i\kappa_{+}/2$, see \cref{fig:re-im-three-R}. Since $\kappa_{+}<\kappa_{-}$, these modes have $\beta\simeq(\kappa_{+}/2\kappa_{-})<1/2$, and SCC conjecture is respected, just like near-extremal KdS BHs~\cite{Dias:2018ynt}. This result is also consistent with that of \cite{Hod:2018lmi}.

\section{Conclusion}
In this work, we have presented an alternative formulation regarding the separability of gravitational and electromagnetic perturbations of KNdS BHs, thereby obtaining a set of decoupled radial and angular equations. An interesting feature of this formulation is that it admits a straightforward $\Lambda\rightarrow 0$ limit, providing a direct connection with the corresponding perturbation equations for KN BHs \cite{Hintz:2026jiw}. Using these equations, we have computed the QNM spectrum through two independent numerical approaches: the spectral method and the direct integration method. The excellent agreement between these two methods provides an important consistency check and establishes the reliability of our numerical results. Our analysis provides a detailed characterization of the QNM spectrum of KNdS BHs, with particular emphasis on the photon sphere and NE mode families. We have explored their dependence on the BH parameters and have identified interesting features associated with the KNdS BH, resulting from the interplay between rotation and electric charge. In particular, the behavior of the real and imaginary parts of the QNM frequencies reveals a non-trivial dependence on the azimuthal quantum number $m$, leading to a Zeeman-like splitting, as well as the existence of fastest decaying modes in the negative $m$ sector. Furthermore, the recovery of the corresponding spectrum of RNdS BH in the non-rotating limit provides another additional consistency check for our results.

One of the central motivations of this study is to understand how the QNM spectrum of KNdS BHs influences the validity of the SCC conjecture. Our results indicate that, unlike the RNdS case, where all the different QNM families--including the NE and dS modes--are essential in determining the fate of SCC conjecture, the photon sphere and the NE modes predominantly govern its validity for KNdS BHs. In fact, we could determine two critical values $\Theta_{\rm c}$ and $\mathcal{R}_{\rm c}$, such that for $\mathcal{R}<\mathcal{R}_{\rm c}$, the competition between the decay of perturbations and the blueshift at the Cauchy horizon is controlled primarily by the photon sphere modes, while for $\mathcal{R}>\mathcal{R}_{\rm c}$, the NE modes dominate the QNM spectrum for KNdS BHs. In both of these cases, for $\Theta<\Theta_{\rm c}$, the SCC conjecture is violated, while otherwise, it is respected. This is consistent with the two extremities, the RNdS BH violates the SCC conjecture, while the Kerr-dS BH respects it. 

\begin{acknowledgments}
The research of S.C. is supported by MATRICS (MTR/\allowbreak 2023/\allowbreak 000049) and Core Research Grants (CRG/\allowbreak 2023/\allowbreak 000934) from SERB, ANRF, Government of India. S.G. acknowledges support of ANRF through NPDF fellowship (PDF/\allowbreak 2025/\allowbreak 004180). S.M.D. acknowledges financial support from ANRF Core Research Grants (CRG/2023/000934).
\end{acknowledgments}
\par\vfill

\onecolumngrid

\renewcommand{\topfraction}{0.9}
\renewcommand{\bottomfraction}{0.75}
\renewcommand{\textfraction}{0.07}
\renewcommand{\floatpagefraction}{0.7}
\setcounter{topnumber}{3}
\setcounter{totalnumber}{4}
\makeatletter
\setlength{\@fptop}{0pt}
\setlength{\@fpsep}{12pt plus 4pt}
\setlength{\@fpbot}{0pt plus 1fil}
\makeatother

\appendix

\section{Separability of the Perturbation Equations}\label{Separability}

\subsection{Steps to separability}
The gravitational and electromagnetic perturbations of a KNdS BH are inherently coupled. Their separation into radial and angular parts therefore requires: (a) Fourier decomposition of the perturbations in $t$ and $\phi$, owing to the stationarity and axisymmetry of the KNdS BH, and (b) a suitable transformation of the perturbation variables. In what follows, we present the linearized Newman-Penrose (NP) equations adapted in \cite{Hintz:2026jiw, Yoshida:2026yly, Zhao:2026gen} and then choose an appropriate algebraic separation matrix, which is key to separating the perturbation equations into radial and angular parts, with a direct limit to that of the KN BH. 

\subsection{The perturbing variables}
Owing to the stationary and axi-symmetry of the background spacetime, the perturbations can be decomposed as, $e^{- i\omega t+im\phi}$, with integer $m$ and complex $\omega$. The linearized NP equations for the Weyl scalars, the Maxwell scalars and the spin coefficients then yield a system of first-order differential equations in $(r,\theta)$, which splits into two sectors, $\varsigma=\pm1$, of positive and negative spin weight. After a rescaling and then taking a linear combination of the perturbation variables, as for the $\Lambda=0$ case \cite{Hintz:2026jiw}, each sector is described by two-component fields $\bm{y}_{\pm}^{(\varsigma)}(r,\theta)$, satisfying the following equations,
\begin{align}
\left(\mathcal{D}\bm{I}-\bm{A}\right)\bm{y}_{+}^{(\varsigma)}
&=\bm{\mathcal{F}}^{(\varsigma)}\,\bm{y}_{-}^{(\varsigma)}~,
\nonumber
\\
\left(\bar{\mathcal{D}}\bm{I}+\bm{A}\right)\bm{y}_{-}^{(\varsigma)}&=\operatorname{adj.}\bm{\mathcal{F}}^{(\varsigma)}\,\bm{y}_{+}^{(\varsigma)}~,
\label{eq:chiral}
\end{align}
where all the boldface symbols denote matrices, $\bm{I}$ is the $(2\times 2)$ identity matrix, and each of the quantities introduced above has the following definitions,
\begin{align}
&\mathcal{D}\equiv\sqrt{\Delta_r}\,\partial_r-i\sqrt{\Delta_{\theta}}\partial_\theta\,;
\quad
\bar{\mathcal{D}}\equiv\sqrt{\Delta_{r}}\,\partial_{r}+i\sqrt{\Delta_{\theta}}\partial_{\theta}\,,
\nonumber
\\
&\bm{\mathcal{F}}^{(\varsigma)}=c^{(\varsigma)}\bm{I}-\bm{B}^{(\varsigma)}\,; 
\quad
c^{(\varsigma)}=\frac{i\varsigma K}{\sqrt{\Delta_r}}+\frac{3\sqrt{\Delta_r}}{\bar{\rho}^{*}}-\frac{3\Delta_{r}'}{4\sqrt{\Delta_r}}\,,
\nonumber 
\\
&\bm{A}=
\begin{pmatrix}
0&-\frac{iQ\bar{\rho}^{*}}{\bar{\rho}^{2}}\\ 
\frac{iQ\bar{\rho}}{(\bar{\rho}^{*})^{2}} & 0
\end{pmatrix}, \quad
\bm{B}^{(\varsigma)}=i
\begin{pmatrix}
b^{(\varsigma)}&\frac{3(M\bar{\rho}-Q^{2})}{Q\bar{\rho}}\\ 
\frac{3Q\bar{\rho}}{(\bar{\rho}^{*})^{2}}&-b^{(\varsigma)}
\end{pmatrix}, 
\nonumber\\
&b^{(\varsigma)}=-\varsigma W-\frac{3\cos\theta\left\{1+\alpha\cos(2\theta)\right\}}{2\sqrt{\Delta_\theta}\sin\theta}+\frac{3ia\sqrt{\Delta_\theta} \sin \theta}{\bar{\rho}^{*}}\,.
\label{eq:AB}
\end{align}
Here $\bar{\rho}=r+ia\cos\theta$ and $\bar{\rho}^{*}$ is the complex conjugate of $\bar{\rho}$, such that $\bar{\rho}\bar{\rho}^{*}=\rho^{2}$, as well as $W\equiv(1+\a)(m-a\omega\sin^2\theta)/(\sqrt{\Delta_\theta}\sin\theta)$ and $K\equiv(1+\a)\{am-(r^2+a^2)\omega\}$. The matrix $\bm{A}$ is Hermitian and $\bm B^{(\varsigma)}$ is traceless, such that $\operatorname{adj}\bm{\mathcal{F}}^{(\varsigma)}=c^{(\varsigma)}\bm{I}+\bm{B}^{(\varsigma)}$. Using these definitions, we can introduce the following variables in terms of the rescaled perturbations of the Weyl scalars, $\widehat\Psi_{j}$, and of the following perturbed spin coefficients, $\left(\widehat\kappa,\widehat\sigma,\widehat\nu,\widehat\lambda\right)$, such that \cite{Hintz:2026jiw, Yoshida:2026yly},
\begin{align}
\bm y_\pm^{(+)}
&=\Delta_r^{5/4}\big({\sqrt{\Delta_\theta}\sin\theta}\big)^{1/2}
\begin{pmatrix}
\widehat \Psi_0\pm i\widehat\Psi_1/\sqrt{\Delta_r}
\\
-2Q(\widehat\kappa\pm i\widehat\sigma/\sqrt{\Delta_r})
\end{pmatrix}\,,
\label{eq:chiral_fields01}
\\
\bm y_\pm^{(-)}
&=\Delta_r^{-3/4}\big({\sqrt{\Delta_\theta}\sin\theta}\big)^{1/2}
\begin{pmatrix}
\widehat\Psi_4\mp i\sqrt{\Delta_r}\widehat\Psi_3
\\
2Q(\widehat\nu\mp i\sqrt{\Delta_r}\widehat\lambda)
\end{pmatrix}~,
\label{eq:chiral_fields02}
\end{align}
In the $\Lambda\to0$ limit, one has $\sqrt{\Delta_r}\to\sqrt{\Delta_0}$, ${\Delta_\theta}\to1$ and $W\to (m/\sin\theta)-a\omega\sin\theta$, and hence \cref{eq:chiral_fields01,eq:chiral_fields02} will coincide with the chiral system of \cite{Hintz:2026jiw} for the KN BH.

\subsection{The algebraic separation matrix}
As evident from \cref{eq:chiral}, the perturbation fields in $\bm{y}_{\pm}^{(\varsigma)}(r,\theta)$ are coupled. To separate the equations into radial and angular sectors, we look for a $2\times2$ matrix $ \bm{X}(r,\theta)$, independent of $\omega$, $m$ and $\varsigma$, with the following properties:
\begin{align}
\text{(T1)}\quad&\det \bm X=1~,
\label{eq:detX}
\\
\text{(T2)}\quad& \mathcal{D}\bm{X}=\bm{A}\bm{X}+\bm{X}\left\{\alpha_r(r)+\alpha_\theta(\theta)\right\}\bm \sigma_2~,
\label{eq:Xeq}
\\
\text{(T3)}\quad& \bm C^{(\varsigma)}\equiv- \bm X^{-1}\bm{\mathcal F}^{(\varsigma)}(\bm X^\dagger)^{-1}
\nonumber
\\
&\quad= R_0^{(\varsigma)}(r)\,\bm I+iR_1^{(\varsigma)}(r)\, \bm \sigma_2
\nonumber
\\
&\qquad+i\Theta_1^{(\varsigma)}(\theta)\,\bm \sigma_1+i\Theta_0^{(\varsigma)}(\theta)\,\bm \sigma_3~.
\label{eq:T}
\end{align}
Here $\bm \sigma_{1,2,3}$ are the Pauli matrices, $\alpha_{r}(r)$ and $\alpha_{\theta}(\theta)$ are real functions, $R_{0}^{(\varsigma)}$ and $R_{1}^{(\varsigma)}$ depend only on the radial coordinate $r$, while $\Theta_0^{(\varsigma)}$ and $\Theta_1^{(\varsigma)}$ depend only on $\theta$. All of these also depend on the frequency $\omega$ and the azimuthal number $m$. In matrix form, the quantity $\bm C^{(\varsigma)}$ has the following expression,
\begin{align}
\bm C^{(\varsigma)}&=\bm R^{(\varsigma)}+i\bm \Theta^{(\varsigma)}\,,
\quad 
\textrm{adj.}\bm C^{(\varsigma)}=(\bm R^{(\varsigma)})^{T}-i\bm \Theta^{(\varsigma)}
\nonumber
\\
\bm R^{(\varsigma)}&=
\begin{pmatrix}R_0^{(\varsigma)}&R_1^{(\varsigma)}
\\
-R_1^{(\varsigma)}&R_0^{(\varsigma)}
\end{pmatrix}\,,
\quad
\bm \Theta^{(\varsigma)}=
\begin{pmatrix}
\Theta_0^{(\varsigma)}&\Theta_1^{(\varsigma)}
\\ 
\Theta_1^{(\varsigma)}&-\Theta_0^{(\varsigma)}
\end{pmatrix}\,.
\label{eq:Rmat}
\end{align}
Among the three conditions satisfied by the separability matrix ${\bm X}$, the condition (T1) in \cref{eq:detX} fixes the normalization of $\bm X$ up to a sign. 
Condition (T2) in \cref{eq:Xeq} makes $\bm X$ an integrating factor for the operator $(\mathcal{D}-\bm A)$, up to a residual $\{\alpha_{r}(r)+\alpha_{\theta}(\theta)\}\bm \sigma_2$, which vanishes for $\Lambda=0$~\cite{Hintz:2026jiw}. Finally, the condition (T3) in \cref{eq:T} separates the perturbation equations into radial and angular parts. This yields, \cref{eq:Firstorder_radial,eq:Firstorder_angular} in the main text. Also the explicit form of various radial and angular functions appearing above can be found in the main text as well as in Appendix~\ref{App:exp}.

The first order radial and angular equations in \cref{eq:Firstorder_radial,eq:Firstorder_angular} of the main text can be converted to second-order decoupled ordinary differential equations for the radial and angular sector, which read,
\begin{align}
&\left[\sqrt{\Delta_r}\,\frac{d}{dr}\pm(\lambda+\alpha_r)\right]
\frac{1}{R_0^{(\varsigma)}\mp iR_1^{(\varsigma)}}
\left[\sqrt{\Delta_r}\,\frac{d}{dr}\mp(\lambda+\alpha_r)\right]\mathcal{R}^{(\varsigma)}_{1,2}
=\left(R_0^{(\varsigma)}\pm iR_1^{(\varsigma)}\right)\mathcal{R}^{(\varsigma)}_{1,2}\,,
\label{KNdSradial}
\\
&\left[\sqrt{\Delta_\theta}\,\frac{d}{d\theta}\pm i(\alpha_\theta-\lambda)\right]
\frac{1}{\Theta_1^{(\varsigma)}\mp i\Theta_0^{(\varsigma)}}
\left[\sqrt{\Delta_\theta}\,\frac{d}{d\theta}\mp i(\alpha_\theta-\lambda)\right]S^{(\varsigma)}_{1,2}
=\left(\Theta_1^{(\varsigma)}\pm i\Theta_0^{(\varsigma)}\right)S^{(\varsigma)}_{1,2}\,,
\label{KNdSangular}
\end{align}
where the upper signs are for $\mathcal{R}^{(\varsigma)}_1$ and $S^{(\varsigma)}_1$, and each operator acts on everything to its right.

The conditions (T1)--(T3) do not fix the separability matrix $\bm X$ uniquely. In particular, if $\bm X$ is one such separability matrix, so is $\bm X\,e^{-\{f(r)+ig(\theta)\}\bm \sigma_2}$ for any real functions $f$ and $g$. This is because, $e^{-\{f(r)+ig(\theta)\}\bm \sigma_2}$ has unit determinant, $\mathcal D\,e^{-\{f+ig\}\bm \sigma_2} =-\{\sqrt{\Delta_r}\,f'+\sqrt{\Delta_\theta}\,g'\}\bm \sigma_2\, e^{-\{f+ig\}\bm \sigma_2}$, and the fact that $\bm \sigma_{2}$ commutes with the radial part of \cref{eq:T} and anticommutes with the angular part. Under this transformation of the separability matrix, we have,
\begin{align}
&\alpha_r\to\alpha_r-\sqrt{\Delta_r}\frac{df}{dr}\,,
\quad
\alpha_\theta\to \alpha_\theta-\sqrt{\Delta_\theta}\frac{dg}{d\theta}\,,
\nonumber
\\
&R_0^{(\varsigma)}\pm i R_1^{(\varsigma)}\to e^{\pm2f(r)}\left(R_0^{(\varsigma)}\pm iR_1^{(\varsigma)}\right)\,,
\nonumber
\\
&\Theta_1^{(\varsigma)}\mp i\Theta_0^{(\varsigma)}\to e^{\mp2ig(\theta)}\left(\Theta_1^{(\varsigma)}\mp i\Theta_0^{(\varsigma)}\right)\,,
\label{eq:freedom}
\end{align}
so that, both the combinations $(R_0^{(\varsigma)})^{2} +(R_1^{(\varsigma)})^{2}$ and $(\Theta_0^{(\varsigma)})^2+(\Theta_1^{(\varsigma)})^2$ are invariant.

Focussing on our case of interest, i.e., for KNdS BH, describing spacetime with mass, electric charge, rotation and positive cosmological constant, we introduce the following quantities, 
\begin{align}
\eta\equiv \frac{2Q^2}{\nu_0+\alpha Q^2+\nu}~,
\qquad
q\equiv\left(\frac{1-\alpha\eta\cos^2\theta}{1+\alpha\eta\,r^2/a^2}\right)^{1/2}~,
\label{Xh:eta}
\end{align}
such that we obtain the following separability matrix, 
\begin{align}
{\bm X}=\frac1{\sqrt\nu}\begin{pmatrix}
\dfrac{iQ\left(q\,\sqrt{\Delta_r}+a\sin\theta\sqrt{\Delta_\theta}/q\right)}{\bar \rho}&\dfrac{q\,(Mr-Q^2)+iaM\cos\theta/q}{\bar\rho}\\[4mm]
-\dfrac{q\,(Mr-Q^2)-iaM\cos\theta/q}{\bar\rho^*}&\dfrac{iQ\left(q\,\sqrt{\Delta_r}-a\sin\theta\sqrt{\Delta_\theta}/q\right)}{\bar\rho^*}
\end{pmatrix} \exp(\psi \bm \sigma_2)
\label{eq:ourX}
\end{align}
One can immediately verify that the above separability matrix satisfies all the conditions presented above in \cref{eq:detX,eq:T}. The phase $\psi$, appearing in the above separation matrix $\bm X$ can be expressed as, $\psi=\psi_{r}(r)+i\psi_{\theta}(\theta)$, with the following expressions,
\begin{align}
&e^{2i\psi_\theta}=a^2 \frac{Q\cos\theta-iM\sin\theta}{\sqrt{H_0 H_{\theta}}\,\big(1-\alpha\eta\cos^2\theta\big)}
\Big[Q\cos\theta\Big(1-\alpha-2\alpha\eta+\alpha(2+\eta-\alpha\eta)\cos^2\theta\Big)+iM\sqrt{\Delta_\theta}\,\sin\theta\,\big(1+\alpha\eta\cos^2\theta\big)\Big]\,,
\\
&e^{2\psi_r}=\frac{\mathcal S_0}{\mathcal S_r\,N^{-}_0}\bigg[\nu\,(r^2+a^2)
+\frac{2Q(1-\alpha\eta)}{1+\alpha\eta\,r^2/a^2}\Big(Q\Delta_r-(Mr-Q^2)\sqrt{\Delta_r}\Big)\bigg]\,.
\label{Xh:psidef}
\end{align}
The explicit expressions for several quantities appearing in the above expression for the separability matrix and the phase $\psi$ can be found in Appendix~\ref{App:exp}. Further, the condition (T3), along with the above separation matrix, gives the expressions for the components of $\bm R^{(\varsigma)}$ and $\bm{\Theta}^{(\varsigma)}$ for KNdS BH.

\subsection{Kerr-Newman limit}
As $\Lambda\to 0$, we get back the asymptotically flat KN BH. In particular, with vanishing cosmological constant, we have the following limit for various quantities appearing in the perturbation equations and the separability matrix: $\alpha\to0$, $\nu\to\nu_0$, $\Delta_r\to\Delta_0$, $\Delta_\theta\to1$, $H_{\theta}\to H_{0}$, $\mathcal S_r\to\mathcal S_0$, $q\to1$ and $\exp(\psi\,\bm\sigma_2)\to\bm I$. As a consequence, it turns out that the separability matrix $\bm X$ in \cref{eq:ourX} becomes the integrating factor of~\cite{Hintz:2026jiw} in its unit-determinant form. Also the functions, $\alpha_{r}(r)$ and $\alpha_{\theta}(\theta)$ vanish, and the entries of the ($2\times 2$) matrix $\bm R^{(\varsigma)}$ and $\bm\Theta^{(\varsigma)}$ reduce to the respective KN expressions of Refs.~\cite{Hintz:2026jiw, Destounis:2026ljd}.

\section{The Kerr--Newman--de~Sitter Parameter Space}\label{sec:AppA}

In this section, we present the parameter space of the KNdS solution of the Einstein--Maxwell equations, with cosmological constant $\Lambda$, mass $M$, electric charge $Q$ and angular momentum $J=aM$, that is relevant for this work. We also depict the regions that are potentially interesting but have not been covered in the present analysis.  

Even though the metric written in \cref{eq:metricKNdS} is valid for the asymptotically anti-de Sitter case as well, we will restrict ourselves to the positive cosmological constant case, such that the BH spacetime has a cosmological horizon, located at $r_{\rm c}$, in addition to the event horizon at $r_{+}$ and the inner (Cauchy) horizon at $r_{-}$. We will also assume that $0<r_{-}<r_{+}<r_{\rm c}$. We also take this opportunity to introduce some terminology that is commonly used in the literature (see, for example~\cite{Park:2023lfc} and also see Appendix B of \cite{Furugori:2026bqu} for a detailed discussion and analytical estimates),
\begin{itemize}
    \item $r_{-}=r_{+}$ is termed an extremal BH.
    \item $r_{+}=r_{\rm c}$ is the Nariai limit of the KNdS spacetime.
    \item $r_{+}=r_{\rm c}=r_{-}$ is the critical or ultracold geometry.
\end{itemize} 
In the main text, we will be primarily concerned with the extremal limit. The Nariai and the ultracold limits of the KNdS BH spacetime will be discussed elsewhere.  

At first glance the parameter space appears to be four-dimensional, containing the following parameters $(M,Q,a,\Lambda)$. However, we can reduce it to three dimensions, by properly rescaling all the other parameters by the mass $M$ of the BH, such that our interest is in the following three-dimensional parameter space $\{(a/M),(Q/M),\Lambda M^{2}\}$. To visualize this three-dimensional parameter space and to enforce the condition that we must have a BH with a positive cosmological constant, we have to ensure that $\Delta_{r}$ in \cref{eq:metric_variables_1} has three distinct positive roots. For this purpose, we have to impose the following condition on the discriminant of $\Delta_r$, assuming $\Lambda>0$, which we denote by ${\rm disc.}\Delta_{r}$,
\begin{equation}
{\rm disc.}\Delta_r \equiv -\frac{16\Lambda}{243}\,\mathbb{F} >0\,,
\end{equation}
or, equivalently, $\mathbb{F}<0$, where
\begin{align}
\mathbb{F}&=-81(M^2-a^2-Q^2)+\Lambda^4 a^8(a^2+Q^2)+3\Lambda^3a^4\left(M^2a^2+4a^4+12Q^2a^2+8Q^4\right)
\nonumber
\\
&+9\Lambda^2\left(36M^2Q^2a^2+33M^2a^4+16Q^6+32Q^4a^2+22Q^2a^4+6a^6\right)
\nonumber
\\
&+27\Lambda\left(27M^4-36M^2Q^2-33M^2a^2+8Q^4+12Q^2a^2+4a^4\right)\,.
\end{align}

\begin{figure}[!htbp]
    \centering
    \includegraphics[width=0.48\textwidth]
        {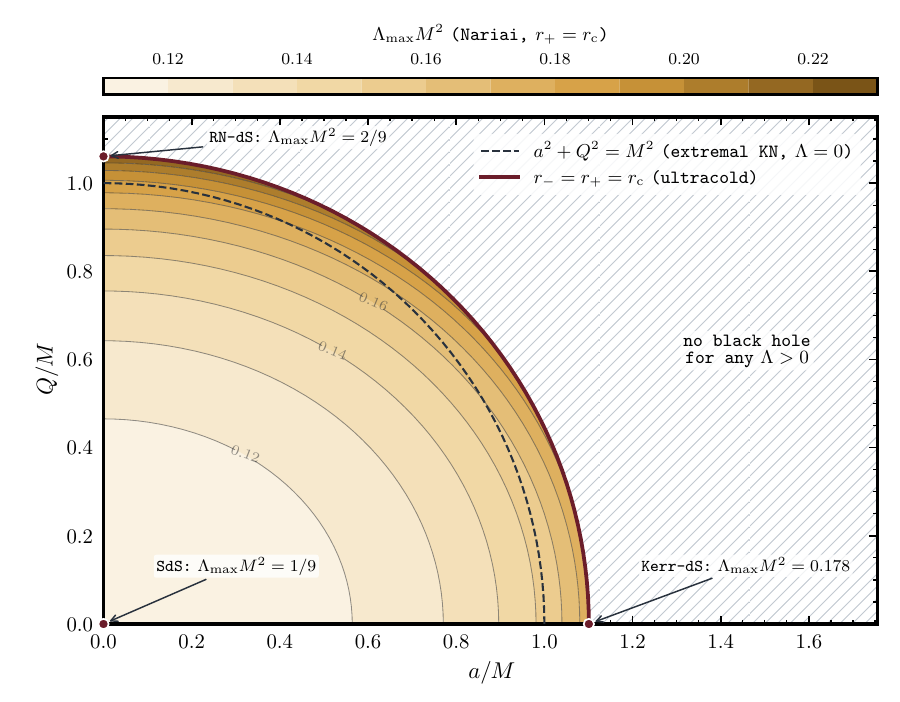}
    \hfill
    \includegraphics[width=0.48\textwidth]
        {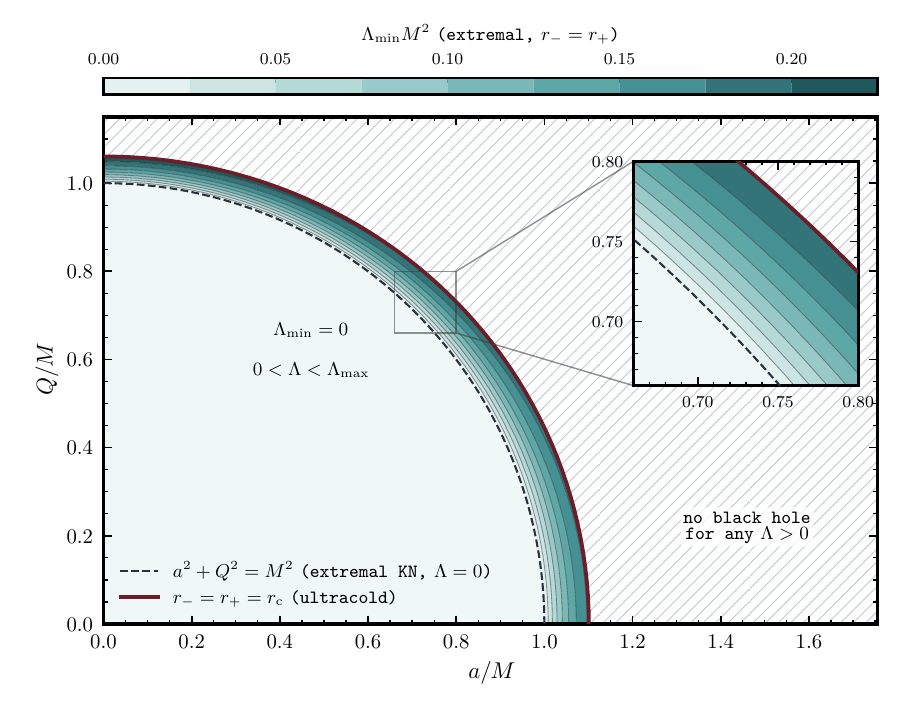}
    \caption{\textbf{Allowed parameter space for sub-extremal KNdS BHs with three horizons}, satisfying $r_{-}<r_{+}<r_{\rm c}$, which exist for $\Lambda_{\min}<\Lambda<\Lambda_{\max}$. Dashed: the extremal KN BH, $a^{2}+Q^{2}=M^{2}$; solid: the ultracold limit $r_{-}=r_{+}=r_{\rm c}$, beyond which no BHs exist (hatched). \textbf{Left:} $\Lambda_{\max}M^{2}$, along with the Nariai limit $r_{+}=r_{\rm c}$. \textbf{Right:} $\Lambda_{\min}M^{2}$, along with the extremal limit $r_{-}=r_{+}$, which vanishes inside the circle; the inset magnifies the region near $a=Q$.}
    \label{fig:knds_parameter_space}
\end{figure}

Using the above expression for $\mathbb{F}$, we can infer the following properties of the $\{(a/M),(Q/M),\Lambda M^{2}\}$ parameter space,
\begin{itemize}

    \item For $a=Q=0$, we get $\mathbb{F}=81M^{2}(9\Lambda M^{2}-1)$, so $\mathbb{F}<0$ gives the following bound on the cosmological constant: $0<9\Lambda M^2<1$. This is the standard condition for the Schwarzschild--de Sitter BH to have two distinct horizons, $r_{+}<r_{\rm c}$.

    \item For $a^{2}+Q^{2}>M^{2}$, i.e., for parameters that describe a naked singularity when $\Lambda=0$, a BH with three horizons will exist only for $\Lambda>\Lambda_{\min}>0$. Thus, there are cases where we need the positive cosmological constant to start from a minimum value $\Lambda_{\min}>0$ \cite{Stuchlik:2000pla,Franzen:2020gke,Furugori:2026bqu}. For example, for $a=0$ and $Q=1.02\,M$, three horizons exist only for $0.103<\Lambda M^2<0.196$.

    \item The existence of BHs with three horizons always requires an upper bound $\Lambda_{\max}$ on the cosmological constant, which increases with both $a$ and $Q$ and satisfies the following bounded inequality $(1/9M^{2})\le\Lambda_{\max}\le (2/9M^{2})$. The lower value is for the Schwarzschild--de Sitter BH $(a=0=Q)$, while the upper value is reached for the RNdS BH with $a=0$ and $Q^{2}=(9M^{2}/8)$, for which $r_{-}=r_{+}=r_{\rm c}$.

    \item  For $a^{2}+Q^{2}<M^{2}$, $\Lambda_{\min}=0$ even when $a$ and $Q$ are both non-zero. 

\end{itemize}
Thus, for a KNdS BH to have three distinct horizons, the positive cosmological constant must be bounded from below, as well as from above, i.e., $\Lambda_{\rm min}\leq\Lambda\leq\Lambda_{\rm max}$. We present the dependence of $\Lambda_{\rm min}$ and $\Lambda_{\rm max}$ on the electric charge and rotation in \cref{fig:knds_parameter_space}. For $\Lambda\to0$, on the other hand, the condition on the discriminant of $\Delta_{r}$ becomes $a^{2}+Q^{2}<M^{2}$: the well-known Kerr--Newman bound for a BH with two distinct horizons.

For the numerical computations, it is often useful to consider the spherical polar parametrization of the KNdS BH, since it allows us to easily traverse the parameter space from the RNdS to the KdS limit. Following Ref.~\cite{Davey:2024xvd}, we label the BHs by $(y_{+},\mathcal{R}^{2},\Theta)$, which are defined in \cref{eq:sphpolparam} of the main text, and hence introduce the following auxiliary parameters:  
\begin{align}
L_{0}&=1 + y_+\left\{1 + \mathcal{R}^2 + y_+\left(1 + \mathcal{R}^2 + \mathcal{R}^4\right)\right\}\,, 
\\
L_2 &=\frac{L_0}{2}\left[1+\sqrt{1+\frac{4y_{+}^{2}\mathcal{R}^2\left(1+y_{+}+y_{+}\mathcal{R}^2\right)\sin^2\Theta}{L_0^2}}\right]\,, 
\\
\bar{s}&=\frac{y_{+}\mathcal{R}\sqrt{1 + y_{+}\left(1+\mathcal{R}^{2}\right)}}{\sqrt{L_2}}\,.
\end{align}
In terms of these newly introduced parameters, the BH hairs (mass, angular momentum, electric charge) and the cosmological constant can be expressed as,
\begin{equation}
\Lambda=\frac{3}{L_{2}}\,;
\quad
M=\frac{y_{+}\left(1+y_{+}\right)\left(1 + \mathcal{R}^2\right)\left(1+y_{+}\mathcal{R}^2\right)}{2L_2}\,; 
\quad 
a=\bar{s}\sin\Theta\,;
\quad 
Q=\bar{s}\cos\Theta\,. \label{eq:MaQ}
\end{equation}
We have used these parameters in the numerical results for the QNM frequencies presented in the main text. 

\section{Notations and Expressions for Relevant Quantities in the Separable Equations}\label{App:exp}

In this section, we have collected some useful definitions and expressions of quantities that have been used throughout the main text. To start with, we introduce the following quantities present in the algebraic separation matrix $\bm{X}$, which we have introduced in the main text,
\begin{align}
&\Delta_{r}=\Delta_{0}+\alpha\,\delta\Delta\,;
\quad
\delta\Delta\equiv-\frac{r^{2}}{a^{2}}(r^{2}+a^{2})\,,
\\
&\nu^{2}=\nu_{0}^{2}+\alpha \underbrace{Q^{2}\left[2(M^{2}+Q^{2})+\alpha Q^{2}\right]}_{\delta \nu}\,;
\quad
\nu_{0}\equiv \left(M^{2}-Q^{2}\right)\,,
\\
&\mathcal{S}^{2}_{r}=\mathcal{S}_{0}^{2}+\alpha \underbrace{(r^{2}+a^{2})^{2}\delta \nu+4a^{2}M^{2}Q^{2}\delta \Delta}_{\delta \mathcal{S}}\,;
\quad
\mathcal{S}_{0}^{2}\equiv \nu_{0}^{2}(r^{2}+a^{2})^{2}+4a^{2}M^{2}Q^{2}\Delta_{0}\,,
\\
&\mathcal{Q}_{r}=\mathcal{Q}_{0}-\alpha (r^{2}+a^{2})\left[-2MrQ^{2}-M^{2}(r^{2}-a^{2})+(\alpha+2)Q^{2}(r^{2}+a^{2})\right]\,,
\\
&\mathcal{Q}_{0}\equiv \left\{2Mr-(r^{2}+a^{2})\right\} \left[\left(r^{2}+a^{2}\right)Q^{2}-\left(r^{2}-a^{2}\right)M^{2}\right]+4a^{2}M^{2}\left(Mr-Q^{2}\right)~,
\\
&S_{+}=(r^{2}+a^{2})\nu_{0}+2Q^{2}\Delta_{0}+\alpha Q^{2}\delta \Delta=(r^{2}+a^{2})\nu_{0}+2Q^{2}\Delta_{r}-\alpha Q^{2}\delta \Delta\,;
\quad
S_{-}=2Q(Mr-Q^{2})\sqrt{\Delta_{r}}\,,
\\
&S^{(0)}_{+}=(r^{2}+a^{2})\nu_{0}+2Q^{2}\Delta_{0}\,;
\quad
S^{(0)}_{-}=2Q(Mr-Q^{2})\sqrt{\Delta_{0}}\,,\\
&N^{-}_0=S_+^{(0)}-S_-^{(0)}\,; \quad N^{+}_0=S_+^{(0)}+S_-^{(0)}~,\\
&H_{\theta}=H_{0}+\alpha a^{2}\cos^{2}\theta\left[M^{2}\sin^{2}\theta+(\alpha+2)Q^{2}\right]~; 
\quad
H_{0}\equiv a^{2}\left[M^{2}\sin^{2}\theta+Q^{2}\cos^{2}\theta\right]~,\\
&\mathbb E_\Lambda\equiv(r^2+a^2)\Delta_r'-4r\Delta_r\,;
\qquad
\mathbb {E}_0\equiv(r^2+a^2)\Delta_0'-4r\Delta_0\,.
\end{align}
Using these definitions, we can compute the quantities $\alpha_{r}(r)$ and $\alpha_{\theta}(\theta)$, appearing in \cref{eq:Xeq}, which have the following expressions,
\begin{align}
&\alpha_{r}(r)=\frac{Q}{2M}\left[\frac{\nu\mathcal Q_r}{\mathcal S_r^2}-(1+\alpha)-\sqrt{\frac{\Delta_r}{\Delta_0}}\left(\frac{\nu_0\mathcal Q_0}{\mathcal S_0^2}-1\right)\right]~,
\nonumber
\\
&\alpha_{\theta}=\frac{Q}{2M}\left[1-\sqrt{\Delta_{\theta}}+\alpha+a^{2}\cos^{2}\theta \left\{\frac{\nu}{H_{\theta}}-\frac{\nu_{0}}{H_{0}}\sqrt{\Delta_{\theta}} \right\}+\alpha(2+\alpha)\,a^{2}\cos^{2}\theta\,\frac{\nu}{H_{\theta}}\right]~.
\end{align}
Note that in the $\Lambda \to 0$ limit, both $\alpha_{r}$ and $\alpha_{\theta}$ vanish identically. For the entries of the matrices $\bm{R}^{(\varsigma)}$ and $\bm{\Theta}^{(\varsigma)}$, we have the following expressions for the quantities $\widetilde{R}_{0}^{(\varsigma)}$, $\widetilde{R}_{1}^{(\varsigma)}$, $\widetilde{\Theta}_{0}^{(\varsigma)}$ and $\widetilde{\Theta}_{1}^{(\varsigma)}$, which are given by,
\begin{align}
&\widetilde{R}_{0}^{(\varsigma)}=\frac{3}{4\sqrt{\Delta_{r}}}\Bigg[\frac{S_{-}^{(0)}}{\mathcal{S}_{0}\mathcal{S}_{r}}\left\{ \frac{\nu}{\nu_{0}}\Delta_{r}'S_{-}^{(0)}-\Delta_{0}'S_{-}+2Q\alpha\sqrt{\Delta_{r}}\left[M(a^2-r^2)+2Q^2r\right]\right\}\nonumber\\&+\frac{2S_{+}^{(0)}}{\mathcal{S}_{0}\mathcal{S}_{r}}\left\{-\frac{\nu}{\nu_{0}}Q^{2}\Delta_{0}\Delta_{r}'+2\Delta_{r}\left[M(Mr-Q^{2})\sqrt{\frac{\Delta_{0}}{\Delta_{r}}}-r\nu\right] \right\}\Bigg]
\nonumber
\\
&+\frac{i\varsigma(1+\alpha)}{\sqrt{\Delta_{r}}}\Bigg[-\frac{S_{-}^{(0)}am}{\mathcal{S}_{r}\mathcal{S}_{0}}\left(\frac{\nu_{0}}{\nu}S_{-}^{(0)}-S_{-}\right)
-\frac{\alpha amS_{+}^{(0)}}{\nu\mathcal{S}_{r}\mathcal{S}_{0}}\left\{(r^{2}+a^{2})\delta \nu-2Q^{4}\Delta_{r}+2Q^{2}\nu_{0}\delta \Delta\right\}\Bigg]~,
\\
\widetilde{R}_{1}^{(\varsigma)}&=\frac{3i}{4\mathcal{S}_{0}\mathcal{S}_{r}}\left[S_{-}^{(0)}\left(\frac{\nu\,\mathbb E_\Lambda}{\sqrt{\Delta_r}}-\frac{\nu_0\,\mathbb E_0}{\sqrt{\Delta_0}}\right)+2\alpha QS_{+}^{(0)}\left\{M(a^2-r^2)+2Q^2r\right\}\right]
\nonumber\\
&\quad-\frac{\varsigma(1+\alpha)am}{\sqrt{\Delta_{r}}\,\mathcal{S}_{0}\mathcal{S}_{r}}\left[S_{+}^{(0)}\left(S_{-}-\frac{\nu_{0}}{\nu}S_{-}^{(0)}\right)-\frac{\alpha S_{-}^{(0)}}{\nu}\left\{(r^{2}+a^{2})\delta \nu-2Q^{4}\Delta_{r}+2Q^{2}\nu_{0}\delta \Delta\right\}\right],
\\
&\widetilde{\Theta}_{0}^{(\varsigma)}=\frac{\varsigma(1+\alpha)a^{2}}{\sqrt{H_{0}H_{\theta}}}\Bigg[aQ^{2}\omega\sin\theta\left\{\cos^{2}\theta\left(\frac{1+\alpha}{\sqrt{\Delta_{\theta}}}-\frac{\nu_{0}}{\nu}\right)-\frac{\alpha M^{2}(1+\cos^{2}\theta)}{\nu}\right\}
\nonumber\\
&\qquad+m\left\{M^{2}\sin\theta\left(\frac{\nu_{0}}{\nu}-\frac{1}{\sqrt{\Delta_{\theta}}}\right)+\alpha Q^{2}\left(\frac{M^{2}\sin\theta(1+2\cos^{2}\theta)}{\nu}-\frac{\cos^{2}\theta(1-2\cos^{2}\theta)}{\sin\theta\sqrt{\Delta_{\theta}}}\right)\right\}\Bigg]~,\\
&\widetilde{\Theta}_{1}^{(\varsigma)}=\frac{\varsigma(1+\alpha)a^2MQ\cos\theta}{\sqrt{H_{0}H_{\theta}}}\Bigg[-\frac{a\omega}{\nu}\left\{\sin^{2}\theta\left(\frac{\nu}{\sqrt{\Delta_{\theta}}}-\nu_{0}\right)+\alpha \left(Q^{2}(1+\cos^{2}\theta)+\frac{\nu \sin^{2}\theta}{\sqrt{\Delta_{\theta}}} \right)\right\}
\nonumber\\
&\qquad\qquad+m\left\{\frac{\nu_{0}}{\nu}-\frac{1}{\sqrt{\Delta_{\theta}}}+\alpha\left(\frac{Q^{2}(1+2\cos^{2}\theta)}{\nu}+\frac{1-2\cos^{2}\theta}{\sqrt{\Delta_{\theta}}} \right)\right\}\Bigg].
\end{align}
all of which vanish in the $\Lambda\to 0$ limit. These expressions have been used in the main text, as well as in the numerical analysis. 

\section{Analytic Structure of the Angular Eigenvalue}
\label{sec:lambda}

The angular system of equations, as presented in \cref{eq:Firstorder_angular}, in terms of the angular variables $S_{1,2}$ can be cast in a suggestive form by using the following variables: $U\equiv S_{1}+S_{2}$ and $V\equiv S_{1}-S_{2}$, such that the angular system becomes
\begin{align}
\bm{\mathcal{A}}
\begin{pmatrix}
U
\\
V
\end{pmatrix}
=\lambda
\begin{pmatrix}
U
\\
V
\end{pmatrix}\,,
\qquad
\bm{\mathcal{A}}\equiv
\begin{pmatrix}
\alpha_\theta-\Theta_1^{(\varsigma)} & i\big(\sqrt{\Delta_\theta}\,\partial_\theta-\Theta_0^{(\varsigma)}\big)
\\
i\big(\sqrt{\Delta_\theta}\,\partial_\theta+\Theta_0^{(\varsigma)}\big) & \alpha_\theta+\Theta_1^{(\varsigma)}
\end{pmatrix}\,,
\label{eq:Aop}
\end{align}
where $\lambda$ is an eigenvalue of $\bm{\mathcal{A}}$ on functions regular at both the poles of the sphere. The eigenvalue $\lambda$ does not depend on $\varsigma$: in $\Theta_{0}^{(\varsigma)}$ and $\Theta_{1}^{(\varsigma)}$, the spin label $\varsigma$ as well as the angular dependence on $\theta$ appears such that, under $\varsigma\to-\varsigma$ and $\theta\to\pi-\theta$, we obtain, $\Theta_{0}^{(\varsigma)}(\theta)\to -\Theta_{0}^{(-\varsigma)}(\pi -\theta)$ and $\Theta_{1}^{(\varsigma)}(\theta)\to \Theta_{1}^{(-\varsigma)}(\pi -\theta)$, respectively. As a result, the eigenvector $(U,V)(\theta)$ associated with $\varsigma$ transforms to the following eigenvector: $(U,-V)(\pi-\theta)$ as one transforms to the angular system connected with $-\varsigma$, but with the same eigenvalue $\lambda$. Hence, we set $\varsigma=1$ and expand the eigenvalues in $a\omega$ and $\alpha=(\Lambda a^{2}/3)$ to first order, in the spirit of~\cite{Suzuki:1998vy}.

Consider first the case of KN BH. For $\Lambda=0$ and $M>Q$ it follows that we can expand the differential operator associated with the angular eigenvalue problem as, $\bm{\mathcal{A}}^{\rm KN}=\bm{\mathcal{A}}_{0}^{\rm KN}+a\omega\,\bm{\mathcal{A}}_{\omega}^{\rm KN}$, with 
\begin{align}
\bm{\mathcal{A}}_{0}^{\rm KN}=
\begin{pmatrix}
\frac{3M}{2Q} & i\left\{\partial_{\theta}-\frac{3}{2}\cot \theta-\frac{m}{\sin \theta}\right\}
\\ 
i\left\{\partial_\theta+\frac{3}{2}\cot \theta+\frac{m}{\sin \theta}\right\}&-\frac{3M}{2Q}
\end{pmatrix}\,,
\quad
\bm{\mathcal{A}}_{\omega}^{\rm KN}=
\begin{pmatrix}
\frac{2MQ\cos\theta}{M^{2}-Q^{2}}& i\sin\theta\frac{M^{2}+Q^{2}}{M^{2}-Q^{2}}
\\
-i\sin\theta\frac{M^{2}+Q^{2}}{M^{2}-Q^{2}}&-\frac{2MQ\cos\theta}{M^{2}-Q^{2}}
\end{pmatrix}\,.
\end{align}
Consider now the static limit, $a\omega\to0$, in which case the regular eigenfunctions of $\bm{\mathcal{A}}_{0}^{\rm KN}$ become~\cite{Chakraborty:2026kyf},
\begin{align}
U\propto \sqrt{\sin\theta}\;{}_{2}Y_{\ell m}\,,
\qquad
V\propto \sqrt{\sin\theta}\;{}_{1}Y_{\ell m}\,,
\label{eq:UVstatic}
\end{align}
where ${}_{s}Y_{\ell m}(\theta)$ are the spin-weighted spherical harmonics~\cite{NewmanPenrose1966, Goldberg1967}, normalized as $\int_{0}^{\pi}|{}_{s}Y_{\ell m}|^{2}\sin\theta\,d\theta=1$. Note that the action of the operators $\partial_{\theta}\mp\{(3/2)\cot\theta+(m/\sin\theta)\}$ on $\sqrt{\sin\theta}\,{}_{s}Y_{\ell m}$, are equivalent to $-\eth$ and $-\bar\eth$, and reduces/increases the value of the spin, respectively. Using this eigenfunction and properties of the $\eth$ operator, in the static limit of the KN BH, we obtain the following angular eigenvalue,
\begin{align}
\lambda^{\rm static}_{\rm KN}=b\sqrt{(\ell-1)(\ell+2)+\frac{9M^{2}}{4Q^{2}}}\,,
\label{eq:lambda0}
\end{align}
where $b=\pm 1$ denoting the gravitational and electromagnetic branches, respectively. Note that the above result is valid for $\ell\geq 2$. Equivalently, we can also express the above result as, $3M\pm2Q|\lambda_{\rm KN}^{\rm static}|=3M\pm\sqrt{9M^2+4Q^2(\ell-1)(\ell+2)}$, whose comparison with \cite{Chandrasekhar1983}, immediately provides a check of the separated system of equations.

Let us consider the case of interest, i.e., angular eigenvalues for KNdS BH. The leading order corrections to the angular eingenvalue, as reported in \cref{eq:lambda0}, due to dynamics and non-zero cosmological constant, are obtained from the average values of the diagonal elements of $\bm{\mathcal{A}}_{\omega}\equiv \partial_{a\omega}\bm{\mathcal{A}}|_{\alpha=0=a\omega}$ and of $\bm{\mathcal{A}}_{\alpha}\equiv\partial_{\alpha}\bm{\mathcal{A}}|_{\alpha=0=a\omega}$ in the eigenbasis of $\bm{\mathcal{A}}_{0}^{\rm KN}$. This yields,
\begin{align}
\lambda=\lambda_{\rm KN}^{\rm static}+a\omega\lambda_{\omega}+\alpha\lambda_{\alpha}+\mathcal{O}\big(a^{2}\omega^{2},a\omega\alpha,\alpha^{2}\big)\,,
\label{eq:lambdaseries}
\end{align}
where, the first order corrections to the angular eigenvalues due to dynamics and the existence of a non-zero cosmological constant yields,
\begin{align}
\lambda_{\omega}&=-\frac{m\left[MQ\lambda_{\rm KN}^{\rm static}+(M^2+Q^2)(\ell-1)(\ell+2)+\tfrac92M^2\right]}{\ell(\ell+1)\,(M^{2}-Q^{2})\,\lambda_{\rm KN}^{\rm static}}\,,
\label{eq:lambda1}
\\
\lambda_{\alpha}&=\mathtt{a}_{\ell m}+\frac{\mathtt{b}_{\ell m}}{\lambda_{\rm KN}^{\rm static}}\,,
\label{eq:lambdaalpha}
\end{align}
where, the constants $\mathtt{a}_{\ell m}$ and $\mathtt{b}_{\ell m}$ has the following expressions,
\begin{align}
\mathtt{a}_{\ell m}&=\frac{2MQ\left[\ell(\ell+1)(\ell^2+\ell-3)+m^2(\ell^2+\ell+6)\right]}{\ell(\ell+1)(2\ell-1)(2\ell+3)\,(M^{2}-Q^{2})}-\frac{m^2MQ^3}{\ell(\ell+1)\,(M^{2}-Q^{2})^{2}}\,,
\nonumber
\\
\mathtt{b}_{\ell m}&=\frac{\ell(\ell+1)(\ell^2+\ell-3)(2\ell^2+2\ell+5)+6m^2(\ell^4+2\ell^3+2\ell^2+\ell+6)}{2\ell(\ell+1)(2\ell-1)(2\ell+3)}
\nonumber
\\
&\quad-\frac{9Q^2\left[\ell(\ell+1)-3m^{2}\right]}{2\ell(\ell+1)(2\ell-1)(2\ell+3)\,(M^{2}-Q^{2})}-\frac{m^2(2\ell+1)^2Q^4}{2\ell(\ell+1)\,(M^{2}-Q^{2})^2}\,.
\label{eq:AB2}
\end{align}
\Cref{eq:lambda1,eq:lambdaalpha} holds true for both gravitational and electromagnetic perturbations and also for $\ell=1$. Note that each of the coefficients $\lambda_{\omega}$ and $\lambda_{\alpha}$ are dependent on $\{Q^2/(M^{2}-Q^{2})\}$, which makes the divergence of the above expressions in the $Q\to M$ limit explicit. 

It is worth mentioning that the off diagonal components of the perturbation $\mathcal{A}_{\alpha}$, obtained by expanding $\bm{\mathcal{A}}$ to first order in $\alpha$, contains $(a^{2}/H_{0})$, which explicitly depends on $\theta$ and is not a polynomial in $\cos\theta$. Hence, $\mathcal{A}_{\alpha}$ couples all the $\ell$ modes. Its diagonal element is nevertheless elementary and hence one can obtain a closed form expression, as above. Note that upto linear order in $a\omega$ and $\alpha$, it follows that $\lambda(m,a\omega)=\lambda(-m,-a\omega)$. In particular, $\lambda_{\omega}$ is odd in $m$, while $\lambda_{\alpha}$ is even in $m$. 

\section{The Method of Direct Integration}\label{sec:directint}

In this section, we solve the first-order radial, see~\eqref{eq:Firstorder_radial}, and angular, see~\eqref{eq:Firstorder_angular}, differential equations governing the gravito-electromagnetic perturbations of the KNdS BH, by integrating these equations numerically. From these numerically integrated solutions, we compute the QNM frequencies by imposing appropriate boundary conditions at the horizon \cite{Chandrasekhar:1975zza}. In particular, one starts from two Frobenius series solutions, one centered at the event horizon and the other at the cosmological horizon (as we have a non-zero cosmological constant), and then integrates them to an intermediate point. Matching between these two solutions yields the QNM frequencies~\cite{Kokkotas:1999bd, Berti:2009kk, Konoplya:2011qq, Cardoso:2025npr, Berti:2025hly}. The main difficulty in implementing this numerical method is determining the Frobenius series at both boundaries with sufficient accuracy. The direct integration method, originally developed for single differential equations, was later generalized to coupled systems, where the vanishing of the Wronskian in case of a single differential equation, is replaced by the vanishing of the determinant of a matrix, made out of all the independent solutions~\cite{Pani:2013pma, Pani:2013ija, Pani:2013wsa}. This matrix-valued direct integration method has been used to obtain the QNM frequencies for the coupled gravito-electromagnetic perturbations of the KNdS BH.

For the KNdS BH, the determination of the QNM frequency essentially requires determination of a pair of complex eigenvalues $(\omega,\lambda)$, where $\omega$ is the QNM frequency and $\lambda$ is the separability constant between the angular and the radial system. The determination of the pair $(\lambda,\omega)$ requires the angular system of first order equations to admit a solution that is regular at both $\theta=0$ and $\theta=\pi$, while the radial system must admit a solution which is purely ingoing at the event horizon $r_{+}$ and purely outgoing at the cosmological horizon $r_{\rm c}$.

For these series solutions to be convergent at the boundary, we notice that at both $(r_{+},r_{\rm c})$, $\Delta_{r}$ has a simple root. Hence, $\sqrt{\Delta_{r}}$, appearing in the coefficients of \cref{eq:Firstorder_radial}, results in the introduction of a branch point at both boundaries, due to which the solutions cannot be expanded around those points. We remove this branch point by performing the following change of variable $|r-r_{\rm H}|=y^{2}$, which makes $\sqrt{\Delta_r}$ analytic in $y$ and the radial equations can be cast in the following form
\begin{equation}
y\,\partial_{y}\bm{\mathcal{R}}=\widetilde{\bm{A}}\,\bm{\mathcal{R}}\,,
\qquad 
\widetilde{\bm{A}}=\sum_{k\ge0}^{n}\bm{\mathbb{A}}_{k}\,y^{k}\,,
\label{eq:frob-system}
\end{equation}
where, $\widetilde{\bm A}$, as well as $\bm{\mathbb{A}}_{k}$ are $2\times2$ matrices that are analytic at $y=0$. For computational purposes, we truncate the series around regular singular points at order $n$, whose exact value will be discussed later. The above describes a well-known general form of a system of equations, and the solution can be written as \cite{coddington1955theory,wasow_asymptotic_2018,Teschl2012}:
\begin{equation}
\bm{\mathcal{R}}=y^{\mu}\sum_{k\ge0}^{n}\bm{x}_{k}\,y^{k}\,,
\label{eq:frob-series}
\end{equation}
where $\mu$ is determined from the eigenvalues of $\bm{\mathbb{A}}_{0}\equiv \widetilde{\bm{A}}(y=0)$, representing the correct exponent of the ingoing solution at $r_{+}$ and the outgoing solution at $r_{\rm c}$. With $\mu$ fixed, the coefficients $\bm{x}_{k}$ can be obtained using the following recursion relation:
\begin{equation}
\bigl[(\mu+k)\,\bm{I}-\bm{\mathbb{A}}_{0}\bigr]\,\bm{x}_{k}=\sum_{j=1}^{k}\bm{\mathbb{A}}_{j}\,\bm{x}_{k-j}\,,
\qquad 
k\ge1\,,
\label{eq:frob-recursion}
\end{equation}
The angular system is also solved in the same way by expanding around each pole of the sphere.

Solving the recursion relation~\eqref{eq:frob-recursion} requires the Taylor series coefficients $\bm{\mathbb{A}}_{j}$ up to the truncation order $n$ of the original first-order radial differential equation. Since $\widetilde{\bm A}$ has a non-trivial closed-form expression containing several nested square roots, we compute these matrices $\bm{\mathbb{A}}_{j}$ numerically from Cauchy's integral formula, $\bm{\mathbb{A}}_{k}=(2\pi\mathrm{i})^{-1}\oint \widetilde{\bm A}(y)\,y^{-k-1}\,\mathrm{d}y$. To this end, we sample $\widetilde{\bm A}$ at $N$ equally spaced points, $y_{j}=\rho\,e^{2\pi\mathrm{i} j/N}$, where the $N$ points are taken at $j=0,\cdots,N-1$, located on a circle of radius $\rho$ inside the domain of analyticity of $\widetilde{\bm A}$, which yields,
\begin{equation}
\bm{\mathbb{A}}_k=\frac{1}{N\rho^{k}}\sum_{j=0}^{N-1}\widetilde{\bm A}(y=y_{j})\,e^{-2\pi\mathrm{i}jk/N}\,,
\qquad 
k=0,\cdots,n\,.
\label{eq:fft}
\end{equation}
This represents a sum of discrete Fourier transforms, which we evaluate with the fast Fourier transform~\cite{Cooley:1965zz} method, as implemented in \texttt{numpy.fft.fft}~\cite{Harris:2020xlr}. 

The radius of convergence $\rho_{\rm c}$ is set by the distance from the expansion point to the nearest singularity of the coefficients in the radial differential equation. Since $|r-r_{\rm H}|=y^{2}$, it follows that $\rho_{\rm c}$ is the square root of that distance in $r$ coordinates. For the system of first-order radial equations, as in \cref{eq:Firstorder_radial}, the singularities are located at the other zeros of $\Delta_{r}$, the zeros of $\mathcal S_{r}^{2}$ and of $N^{-}_0N^{+}_0$, and at the branch points $r_{\pm}^{(0)}=M\pm\sqrt{M^2-a^2-Q^2}$ of $\sqrt{\Delta_{0}}$, which enter through  $R_{0}^{(\varsigma)}$ and $R_{1}^{(\varsigma)}$. Since $\Delta_{0}-\Delta_{r}=(\Lambda/3)r^2(r^2+a^2)>0$, the branch point $r_{+}^{(0)}$ lies just inside the event horizon, at a distance of
\begin{equation}
d\equiv r_{+}-r_{+}^{(0)}\simeq\frac{\Lambda}{3}\,
\frac{r^2(r^2+a^2)}{\Delta_0'(r)}\Big|_{r=r_+^{(0)}}=\mathcal{O}(\Lambda)\,.
\label{eq:dsing}
\end{equation}
Thus, for small $\Lambda$, the series around $r_{+}$ converges close to the horizon. We truncate the series around regular singular points at order $n=40$ in general, or $n=60$ near extremality, where the inner horizon $r_{-}$ approaches $r_{+}$, which limits the radius of convergence. Similarly, for the fast Fourier transform, we use  $N=128$ points ($256$ near extremality), and we choose the radius of the circle $\rho=(\rho_{\rm c}/\sqrt{2})$ to stay well within the radius of convergence. 
From their starting points $r_{+/\textrm{c}}$, both of the solutions are integrated to a matching radius using the eighth-order Dormand--Prince method~\cite{Hairer1993-bz} as implemented in \texttt{scipy.integrate.solve\_ivp} (method \texttt{DOP853})~\cite{Virtanen:2019joe}, with relative and absolute tolerances of $10^{-9}$ and $10^{-14}$.

Having discussed about the series solution and its convergence, we now discuss the determination of the QNM frequencies. For this purpose, we set the Wronskian of the system of equations, computed from the event horizon and the cosmological horizon, respectively, to be vanishing, thereby ensuring linear dependency of the solutions. The angular system of equations could be approached using the same techniques, where we will have two regular solutions at $\theta=0$ and at $\theta=\pi$ that are integrated to an intermediate point, say, e.g., $\theta=\pi/2$, and the vanishing of the Wronskian between these two solutions, along with the vanishing of the radial Wronskian will give the desired eigenvalues. Numerically, these eigenvalues are obtained by solving the resulting two-parameter eigenvalue problem by two nested root searches with M\"uller's method~\cite{Muller1956-kj,Press:2007}. In this case, for a given guess value of the QNM frequency $\omega$, the inner search finds $\lambda(\omega)$ as a root of the angular Wronskian and the outer search then finds the QNM frequency as a root of the radial Wronskian. A root is accepted only after checks that it is a genuine zero of the radial Wronskian and that the iteration returns to it when restarted with a slightly shifted initial guess value of the QNM frequency $\omega$.

The initial guesses are generated without any external input. We first compute the modes of the non-rotating RNdS BH at $\Theta=0$ with the same method, and then follow each mode in small steps in $\Theta$, using the modes at the previous values of $\Theta$ to predict the next one. A step is accepted only if the converged mode lies close to this prediction. If the mode shifts significantly from the prediction, the step is refined to guard against jumping to a neighboring mode. These results are then verified to be insensitive to the number of series terms, the size of the fast Fourier transform, and the integrator tolerance before accepting them. 

\section{A Spectral Numerical Recipe}\label{sec:numericspectral}

We apply a variant of the Chebyshev pseudospectral collocation method~\cite{Boyd:2001, Trefethen:2000, Cardoso:2013pza, Dias:2015nua, Dias:2015wqa, Jansen:2017oag} to the set of first order separated radial and angular equations presented in \cref{eq:Firstorder_radial,eq:Firstorder_angular}; the results have been shown in the main text and also in the appendices, and its leading digits have been matched with the direct integration method of Appendix~\ref{sec:directint}. Throughout this method, we set the length scale of the problem by fixing $r_{\rm c}=1$ and solve for the two-parameter eigenvalue problem, with the radial equations on the domain $r_{+}<r<r_{\rm c}$ and the angular equations on the region $0<\theta<\pi$, sharing the frequency $\omega$ and the separation constant $\lambda$. 
Further, imposing regularity at the future horizons it is possible to come up with two new variables, $\tilde{x}$ and $\tilde{y}$, that are smooth everywhere in $r\in [r_{+},r_{\rm c}]$. For the angular equations, near the poles $\theta=0$ and $\theta=\pi$, the regular angular solutions scale as powers of $\theta$ and $(\pi-\theta)$, with exponents that depend on $m$ and $\varsigma$. 
In terms of $x=\cos\theta\in[-1,1]$, we factor out the corresponding powers of $(1-x)$ and $(1+x)$ and solve for smooth functions $f_\pm(x)$.
We then discretize both sets of equations on Chebyshev-Gauss-Lobatto collocation grids, with the endpoints included in our computational domain~\cite{Trefethen:2000, Boyd:2001}. Closer to extremality ($\mathcal R\geq 0.99$), we transform to a new radial coordinate, $s=\ln[(r-r_-)/(r_+-r_-)]\,/\,\ln[(r_{\rm c}-r_-)/(r_+-r_-)]\in[0,1]$, and place the Chebyshev-Gauss-Lobatto points in $s$; the radial derivatives are multiplied by the Jacobian, $(d/dr)=(ds/dr)\,(d/ds)$. This concentrates the grid points near the event horizon~\cite{Capuano:2026tjy, Zhou:2025xta, Hintz:2021vfl}. Because the endpoints are regular singular points of the differential operators, the equations themselves degenerate into regularity conditions at the boundaries, so no additional boundary conditions need to be imposed. In this setup, enforcing the smoothness of the regularized eigenfunctions is sufficient to obtain the QNM solutions~\cite{Cardoso:2013pza, Dias:2015nua, Dias:2015wqa, Jansen:2017oag}. We therefore end up with a discretized system of separated equations that are linear in, but coupled through, both eigenvalues $\omega$ and $\lambda$,
\begin{align}
\left(\bm{M}_{0}+\omega \bm{M}_{\omega}+\lambda \bm{M}_{\lambda}\right)\,\bar{\bm{x}}&=0\,;
\qquad 
\bar{\bm x}=\left(\tilde{x}(r_j), \tilde{y}(r_j)\right)^{\mathsf{T}} \in \mathbb{C}^{2N_{r}}\,, \label{eq:paper-pencils-radial}
\\
\left(\bm{A}_{0}+\omega \bm{A}_{\omega}+\lambda \bm{A}_\lambda\right)\,\bar{\bm y}&=0\,; 
\qquad 
\bar{\bm y}=\left(f_+(x_k), f_-(x_k)\right)^{\mathsf{T}} \in \mathbb{C}^{2N_\theta}\,,
\label{eq:paper-pencils}
\end{align}
where $\bar{\bm x}$ and $\bar{\bm y}$ collect the regular radial and angular functions evaluated at the collocation nodes $r_{j}$ and $x_{k}$, and ${\bm M}_{i}\in \mathbb{C}^{2N_r\times 2N_r}$ and ${\bm A}_{i}\in \mathbb{C}^{2N_\theta\times 2N_\theta}$ ($i \in \{0,\omega,\lambda\}$) are the resulting spectral discretization matrices. Throughout, we denote a grid of $N_r$ radial and $N_\theta$ angular collocation points by $N_{r}\times N_{\theta}$. \cref{eq:paper-pencils-radial,eq:paper-pencils} can be combined to form a two-parameter eigenvalue problem~\cite{Atkinson:1972}. For $\bar{\bm z}=\bar{\bm x}\otimes\bar{\bm y}$, the above system is equivalent to,
\begin{equation}
\bm{\Delta}_1\bar{\bm z}=\omega\,\bm{\Delta}_0\bar{\bm z}\,;
\qquad \bm{\Delta}_{2}\bar{\bm z}=\lambda\,\bm{\Delta}_{0}\bar{\bm z}\,,
\label{eq:paper-joint}
\end{equation}
with $\bm{\Delta}_0=\bm{M}_\omega\otimes \bm{A}_{\lambda}-\bm{M}_{\lambda}\otimes \bm{A}_{\omega}$, $\bm{\Delta}_{1}=\bm{M}_{\lambda}\otimes \bm{A}_{0}-\bm{M}_{0}\otimes \bm{A}_{\lambda}$, and $\bm{\Delta}_{2}=\bm{M}_{0}\otimes \bm{A}_{\omega}-\bm{M}_{\omega}\otimes\bm{A}_{0}$~\cite{Atkinson:1972,Plestenjak:2015}.

Using \cref{eq:paper-joint}, we can obtain all the eigenvalue pairs $(\omega,\lambda)$ of the discretized problem at once, without an initial guess, using a generalized eigenvalue solver such as \textsc{Mathematica}'s \texttt{Eigensystem}. However, the resulting Kronecker-product matrices are of size $4N_{r}N_{\theta}\times 4N_{r}N_{\theta}$, scaling with the product of the radial and angular grid sizes. Although in principle a sufficiently large grid can resolve a significant portion of the eigenspectrum directly using extended internal precision, to keep this step computationally feasible we solve \cref{eq:paper-joint} only on coarse grids in machine precision to generate the seed values. Specifically, for each $m=0,\cdots,|\ell|$, we compute this joint spectrum at a small value $\Theta_{\rm s}\ll 1$, which is equivalent to small rotation, on two coarse grids.
The azimuthal number $m$ is an input, since it enters the separated equations through the dependence $e^{im\phi}$ of the perturbations. The multipole number $\ell$, by contrast, does not appear in the separated equations and cannot be assigned a priori. The joint spectrum at fixed $m$ therefore contains the modes of every $\ell\geq|m|$, and $\ell$ is identified only afterwards, through the RNdS limit. The counter-rotating modes, with $\mathrm{Re}\,\omega<0$, are the mirror images $\omega\to-\bar\omega$ of those with $m<0$. We discard the eigenvalues whose $\omega$ differs between the two grids by more than $10^{-3}$ in relative difference. We then compare the remaining eigenvalues with the QNM frequencies of the RNdS BH with the same horizons, which we obtain independently by solving the two decoupled master equations of Ref.~\cite{Stuchlik:2025mjj} using standard spectral collocation for a single-parameter eigenvalue problem. At this small value of rotation (for $\Theta=\Theta_{\rm s}\ll 1$), the QNM frequencies of several slowly decaying modes lie close to those of the corresponding RNdS QNMs. This enables us to assign an angular multipolar number $\ell$, an overtone number $n$, and a branch label $b$ to each identified mode, where we follow the convention
$b=+1$ for gravitational-led and $b=-1$ for electromagnetic-led modes~\cite{Destounis:2026ljd}. The eigenvalues identified in this way serve as seeds for subsequent stages and allow us to track the trajectory traced out by each mode in the complex frequency plane. 

We use an iterative method, the Newton-Raphson algorithm~\cite{Cardoso:2013pza, Dias:2015nua, Dias:2015wqa, Chung:2023wkd,Miguel:2020uln}, to solve \cref{eq:paper-pencils-radial,eq:paper-pencils} for the eigenvalues ($\omega,\lambda$) and the eigenvectors as the unknowns. We set one component of each eigenvector to one to match the number of unknowns to the number of equations~\cite{Dias:2015nua}, and we also set the internal precision to $50$ digits. We stop the iteration when $|\delta\omega|+|\delta\lambda|<10^{-28}$ or if we have completed $N_{\max}=40$ iterations. When initializing a trajectory, the starting eigenvectors are computed by inverse iteration~\cite{Ipsen:1997,Golub:2013}: first, the angular eigenvector is computed using the seed value of $\omega$, with the seed value of $\lambda$ as the shift, and then the radial eigenvector is computed at the resulting $\lambda$, with the seed value of $\omega$ as the shift.

However, we first have to refine the seed values since they were obtained on a coarse grid using machine precision. We refine each seed by Newton--Raphson iteration on \cref{eq:paper-pencils-radial,eq:paper-pencils}, evaluating it on three suitable grids and keeping it if the values of $\omega$ and $\lambda$ on the two finer grids agree to $10^{-6}$. The convergence tests of Appendix~\ref{sm_sec:convergence} show that the modes are well resolved on the two finer grids. We later use the same two grids to track the modes in $\Theta$. The coarsest grid is included to check that the differences between the grids decrease as the resolution increases. Since, unlike in the joint spectrum, the three grids also differ in the number of angular points, this step tests the convergence of $\lambda$ as well, and it discards any candidate whose $\lambda$ shifts with the angular resolution. On each grid, the angular equations are first solved at the seed value of $\omega$, and the eigenvalue $\lambda$ closest to its seed value is kept, together with its eigenvector; the radial equations are then solved at this $\lambda$, and the eigenvalue $\omega$ closest to its seed value is kept, together with its eigenvector.
From each refined seed, we then follow each mode by moving along $\Theta$ in small steps (of, say $\pi/200$), upwards to $\Theta=0.45 \pi$ and down till $\Theta=\pi/200$. At every later point, the iteration starts from the eigenvectors of the previous point. The starting values of $\omega$ and $\lambda$ are extrapolated from the converged values at the previous points of the trajectory. At the second point we use the values at the first point. From the third point on, they are given by polynomial extrapolation through the previous points (the number of points being at most five). At the end of the iteration, we accept a mode if the last correction is below $10^{-20}$ and $\omega$ lies within 10\% of its prediction. The first criterion, looser than the stopping tolerance, accepts an iteration that stalls slightly above $10^{-28}$ and rejects a mode that has not converged; the second ensures that the iteration has not latched on to a neighbouring mode. If either criterion fails, the starting eigenvectors are recomputed by inverse iteration. If inverse iteration also fails, the starting eigenvectors are recomputed from the full angular and radial eigenvalue problems, as done in the refinement step. Lastly, for $\mathcal R > 0.99$, we use a base step size of $\pi/1000$, and for trajectories that exhibit sharp turns or \emph{loops} (see, for example, \cref{fig:intersection-repulsion}), we refine the step size to $\pi/4000$ or smaller. We compute each trajectory independently on two grids (Appendix~\ref{sm_sec:convergence}), and we regard a point as converged where $\omega$ on the two grids agrees to $10^{-5}$ in relative difference. On a laptop with an Apple M5 processor, the joint spectrum for one value of $m$ on the $49\times17$ and $65\times17$ grids takes about 35 CPU minutes, and refining and tracking one mode from $\Theta=\pi/200$ to $0.45\pi$ on grids of up to $101\times49$ points then takes about 15 CPU minutes on average at $\mathcal R=0.95$.

As $\Theta\to0$, a trajectory must end on the Reissner--Nordstr\"om--de Sitter mode from which its labels were taken. When extrapolated to $\Theta=0$ by a polynomial through its eight points of smallest $\Theta$, every fundamental photon sphere trajectory ends on the corresponding RNdS mode to within $\sim 10^{-8}$, which is less than the uncertainty of the extrapolation (see \cref{tab:ps-rnds-limit}). We have tested the solver in two further cases. In order to verify the robustness of our spectral routines, we computed the QNMs of a neutral massless scalar field in the KNdS background. We were able to reproduce the $\ell=m=0$ modes of Ref.~\cite{Davey:2024xvd} at the same $y_+$ and $\mathcal R$. We have verified our entire setup by computing the coupled gravito-electromagnetic QNMs at $a=0.2M$ and $Q=0.4M$: we computed the modes with $\ell=m=2$ for small values of $\Lambda$ and then extrapolated them to $\Lambda=0$. Our results agree with the Kerr--Newman values reported in Ref.~\cite{Destounis:2026ljd} up to $\sim10^{-8}$. 

\section{Salient Features of the QNM Spectrum}\label{sec_sm:features_qnm_specturm}

\subsection{Overtones of the photon sphere modes}

\begin{figure}[!htbp]
    \centering
    \includegraphics[width=0.325\textwidth]{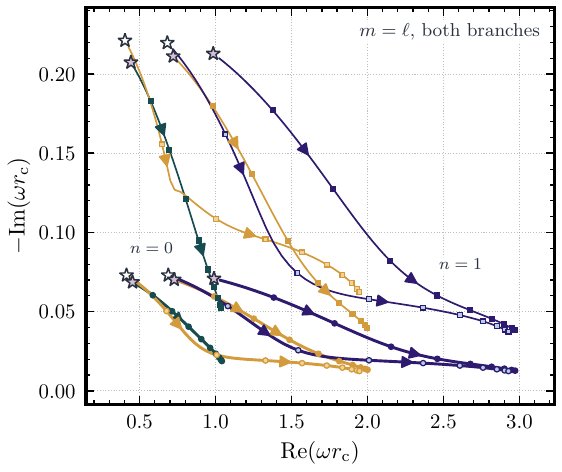}\hfill
    \includegraphics[width=0.325\textwidth]{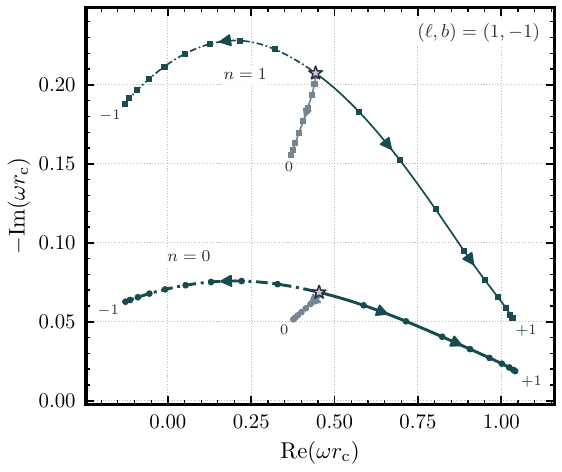}\hfill
    \includegraphics[width=0.325\textwidth]{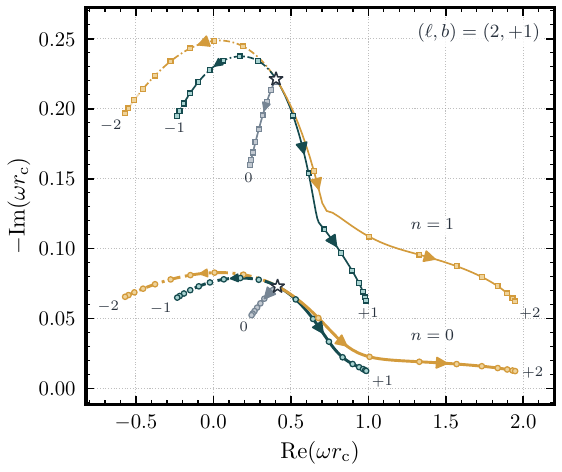}\\[0.1em]
    \includegraphics[width=0.325\textwidth]{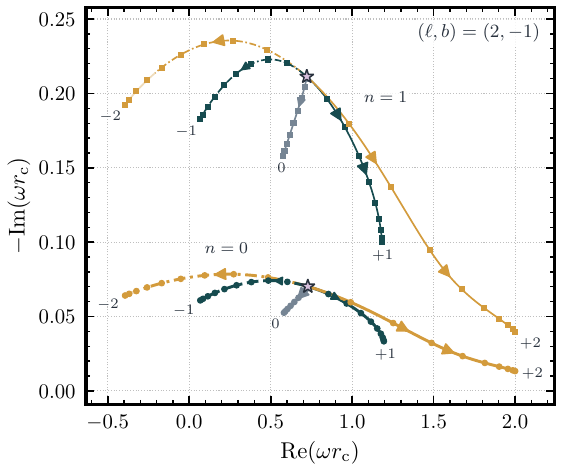}\hfill
    \includegraphics[width=0.325\textwidth]{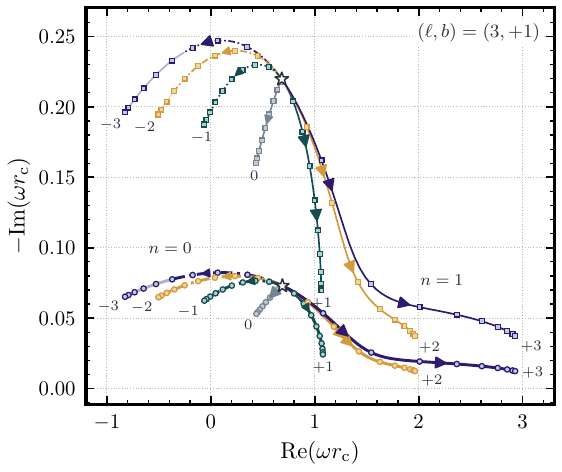}\hfill
    \includegraphics[width=0.325\textwidth]{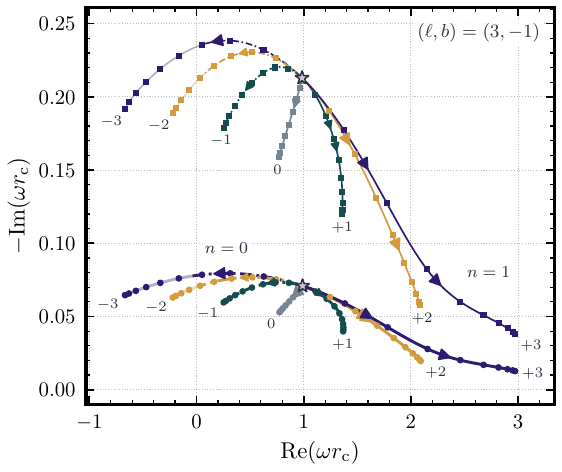}
    \caption{%
\textbf{Fundamental modes ($n=0$, thick curves) and first overtones ($n=1$, thin curves) of the photon sphere family} for $y_+=1/2$ and $\mathcal R=0.95$, arranged as in the main text. Each $n=1$ multiplet fans out from its RNdS mode like the $n=0$ multiplet, but with larger damping.}
    \label{fig:ps_n0_n1_overtone_tracks_by_ell}
\end{figure}

In the main text we focused on the fundamental ($n=0$) photon sphere modes. Here we also present the trajectories of the first overtones ($n=1$) of the same family, with the same BH parameters with $y_{+}=1/2$ and $\mathcal{R}=0.95$, parameterized by $\Theta=\arctan(a/Q)$.  \Cref{fig:ps_n0_n1_overtone_tracks_by_ell} shows both the $n=0$ and $n=1$ photon sphere modes in the complex frequency plane. In every panel the lower curve depicts the fundamental mode, and the upper one is the first overtone. As in the main text, the RNdS limits are marked by stars, filled for $b=-1$ (electromagnetic-led) and hollow for $b=+1$ (gravitational-led). As the rotation increases (equivalently, $\Theta$ varies), the $(2\ell+1)$-fold azimuthal degeneracy of each branch is broken. The $m=\ell$ modes of the electromagnetic-led and gravitational-led branches, which exist for $\ell\geq1$ and $\ell\geq2$, respectively, follow different trajectories (top-left panel of \cref{fig:ps_n0_n1_overtone_tracks_by_ell}). The remaining panels show each multiplet $(\ell,b)$ splitting from the QNMs of the RNdS BH. Selected frequencies of the first overtones are also tabulated in \cref{tab:gem-complete-spectrum} of Appendix~\ref{app:numerical_values_gem_qnms}.

\subsection{Near-extremal modes} 

\begin{figure}[!htbp]
    \centering
    \includegraphics[width=\textwidth]{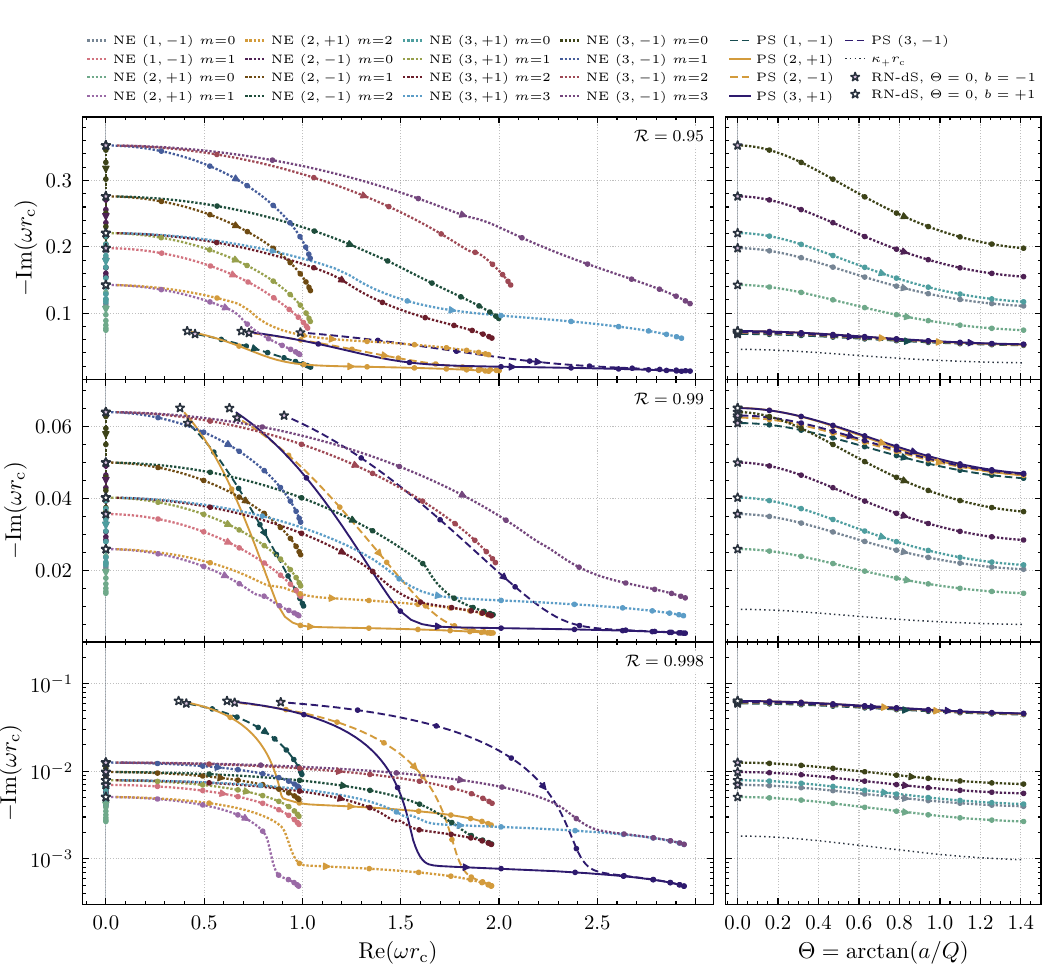}
     \caption{\textbf{Near-extremal (NE) and photon sphere (PS) modes for $\mathcal R=0.95$, $0.99$ and $0.998$} (top to bottom), with $y_+=1/2$. Near-extremal modes are dotted, photon sphere modes are solid for $b=+1$ and dashed for $b=-1$. Stars are the RNdS modes at $\Theta=0$, and arrowheads point towards increasing $\Theta$. For $\mathcal R=0.998$ the damping is shown on a logarithmic scale. \textbf{Left:} the complex frequency plane, with the co-rotating photon sphere modes with $\ell=m$ of both $b=\pm1$ branches; the NE modes with $m=0$ stay on the imaginary axis. \textbf{Right:} $-\mathrm{Im}(\omega r_{\rm c})$ of the NE modes with $m=0$ for each multiplet, together with the photon sphere modes with $m=0$ and the same $(\ell,b)$ label. The curve $-\mathrm{Im}(\omega r_{\rm c})=\kappa_+r_{\rm c}$ is shown by black dots.}
    \label{fig:zdm-tracks-all}
\end{figure}

The NE gravito-electromagnetic QNMs are shown in \cref{fig:zdm-tracks-all} for $\mathcal R=0.95$, $0.99$ and $0.998$, together with the co-rotating photon sphere modes with $\ell=m$. At $\Theta=0$, that is, for the RNdS black hole, the NE modes start on the imaginary axis (marked by stars), with all members of a multiplet originating from the same frequency. As $\Theta$ increases, the members of each multiplet separate out in the complex frequency plane. Those with $m=0$ stay on the imaginary axis, while those with $m\geq1$ acquire real parts, which are close to $m\Omega_{+}$. The interplay between the NE and photon sphere modes depends on how close the KNdS BH is to extremality. Close enough to it, the photon sphere and the NE modes with the same $(\ell,b,m)$ meet, either in an intersection or in an eigenvalue repulsion, which we describe at the end of this section.

Among all the classes of NE modes discussed in \cref{fig:zdm-tracks-all}, the one associated with $\mathcal R=0.998$ is the most interesting one. At $\mathcal R=0.998$, presented by the bottom row of \cref{fig:zdm-tracks-all}, the NE modes start about ten times less damped than the photon sphere modes. Therefore, the least damped mode at small $\Theta$ is the $(2,+1)$ NE mode with $m=2$, and the violation of SCC conjecture is now associated with this mode, rather than a photon sphere mode. For large $\Theta$, all the NE and photon sphere modes with $m\geq1$ have settled on the rungs of the near-horizon ladder, as evident from \cref{fig:zdm-tracks-all}. The NE $(2,+1)$ modes with $m=1$ and $2$ end on $N=0$, $(3,+1)$ mode with $m=3$ on $N=1$ and $(1,-1)$ with $m=1$ on $N=2$. Of the photon sphere modes, $(2,-1)$ and $(3,\pm1)$ end on $N=0$, $(2,+1)$ on $N=2$ after its repulsion, and the dipole $(1,-1)$ on $N=9$.

\subsection{Intersection and repulsion of the modes} 

Among the modes we have tracked, three co-rotating photon sphere modes have a NE partner with the same $(\ell,b,m)$, namely $(2,+1)$ with $m=1$ and $m=2$, and $(3,+1)$ with $m=3$. At large $\Theta$ both members of each pair settle onto the near-horizon ladder, and what happens on the way there depends on how close the black hole is to extremality. For $\mathcal R=0.95$ the two modes never meet. The photon sphere mode stays the less damped of the two over the whole range of $\Theta$, and neither their damping rates nor their real parts cross. Closer to extremality, at $\mathcal R=0.99$, all three pairs meet in what Ref.~\cite{Davey:2024xvd} calls an intersection. The damping rates cross while the real parts are still apart, and afterwards it is the photon sphere mode that ends on the bottom rung, $N=0$, with the near-extremal mode one rung above it, on $N=1$.

\begin{figure}[!htbp]
   \centering
   \includegraphics[width=0.325\textwidth]{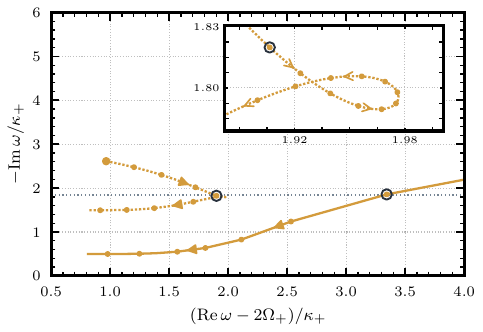}\hfill
   \includegraphics[width=0.325\textwidth]{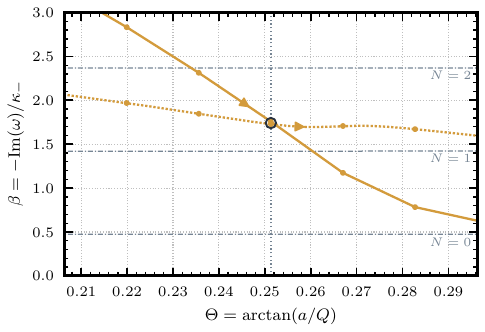}\hfill
   \includegraphics[width=0.325\textwidth]{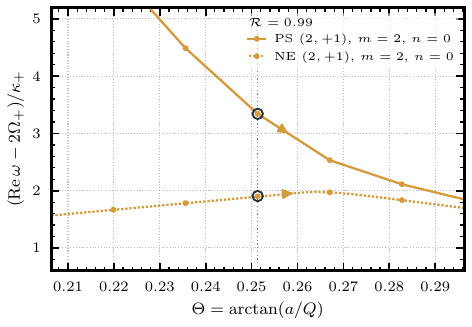}\\[0.25em]
   \includegraphics[width=0.325\textwidth]{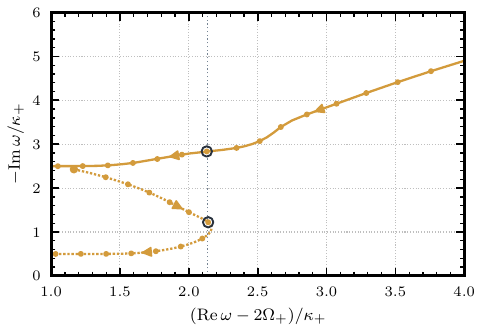}\hfill
   \includegraphics[width=0.325\textwidth]{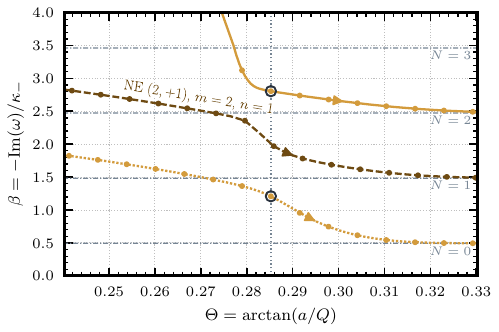}\hfill
   \includegraphics[width=0.325\textwidth]{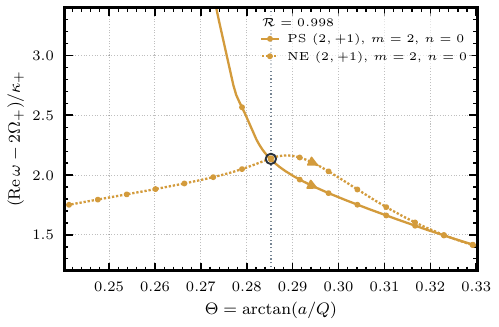}
   \caption{\textbf{From intersection to repulsion}: the co-rotating photon sphere and NE modes with $n=0$ of $(\ell,b)=(2,+1)$ and $m=2$, have been depicted for KNdS BHs with $y_+=1/2$ and $\mathcal R=0.99$ (top) and $\mathcal R=0.998$ (bottom). The arrowheads point towards the direction of increasing $\Theta$. \textbf{Left:} shows the nature of the modes in the complex frequency plane near the encounter (the larger dots mark where the curves start). At $\mathcal R=0.99$ the inset magnifies the small loop the NE mode makes as the photon sphere modes pass by. \textbf{Middle:} the behaviour of the imaginary part of the QNMs with $\Theta$; at $\mathcal R=0.998$ the dashed curve is the NE overtone with $n=1$. \textbf{Right:} describes the real part of the QNM frequency with $\Theta$. Open circles mark the crossing of photon sphere and NE modes. At $\mathcal R=0.99$ the damping rates cross while the real parts differ, an \emph{intersection} in the terminology of Ref.~\cite{Davey:2024xvd}; at $\mathcal R=0.998$ only the real parts cross, an \emph{eigenvalue repulsion}.
   }
   \label{fig:intersection-repulsion}
\end{figure}

\begin{figure}[!htbp]
    \centering
    \includegraphics[width=0.325\textwidth]{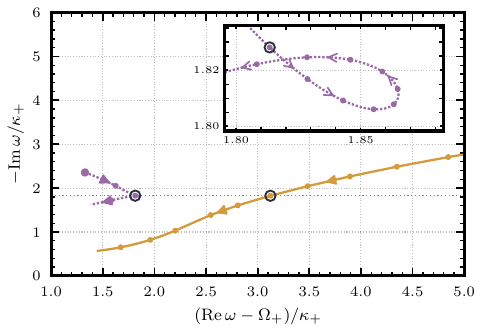}\hfill
    \includegraphics[width=0.325\textwidth]{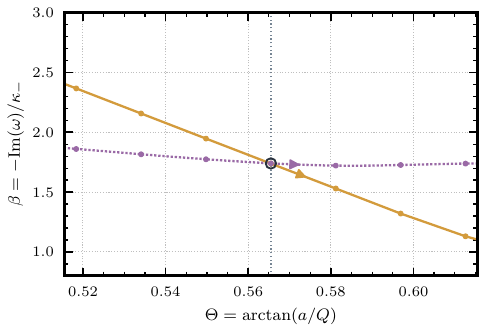}\hfill
    \includegraphics[width=0.325\textwidth]{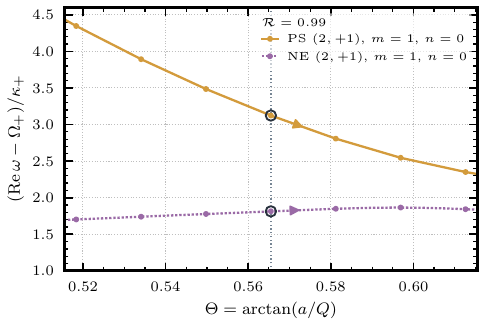}\\[0.25em]
    \includegraphics[width=0.325\textwidth]{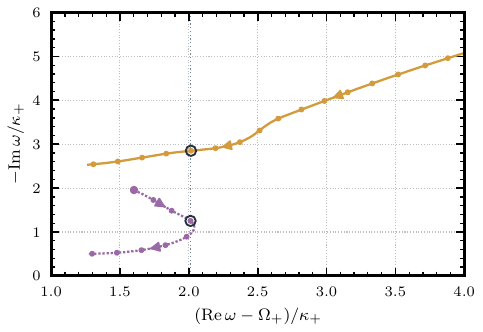}\hfill
    \includegraphics[width=0.325\textwidth]{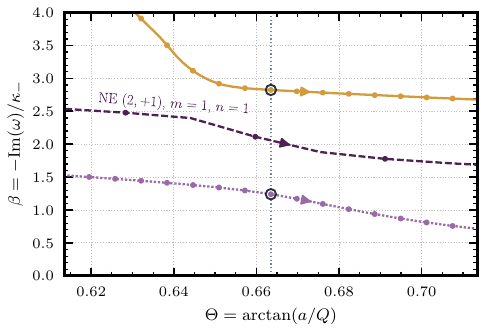}\hfill
    \includegraphics[width=0.325\textwidth]{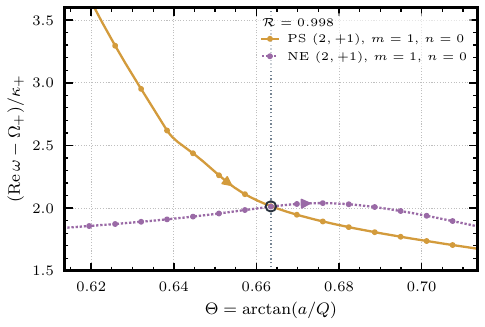}
    \caption{\textbf{The pair $(\ell,b)=(2,+1)$ with $m=1$}, as \cref{fig:intersection-repulsion}: the co-rotating photon sphere (PS, solid) and near-extremal (NE, dotted) modes with $n=0$, for $y_+=1/2$ and $\mathcal R=0.99$ (top) and $\mathcal R=0.998$ (bottom). \textbf{Left:} the complex plane near the encounter; the inset magnifies the small loop of the NE mode at $\mathcal R=0.99$. \textbf{Middle:} $\beta$. \textbf{Right:} the real parts. As in \cref{fig:intersection-repulsion}, open circles mark the PS and NE modes at the value of $\Theta$ where the damping rates cross at $\mathcal R=0.99$, or where the real parts cross at $\mathcal R=0.998$, and dotted lines mark that value of $\Theta$. As for $m=2$, the pair meets as an intersection at $\mathcal R=0.99$, where the damping rates cross at $\Theta=0.565$, and as an eigenvalue repulsion at $\mathcal R=0.998$, where the real parts cross at $\Theta=0.664$ and the NE mode remains the less damped. At $\mathcal R=0.998$ the NE mode ends on $N=0$ and the PS mode on $N=2$, and the dashed curve in the bottom middle panel is the NE overtone with $n=1$, which ends on $N=1$.}
    \label{fig:encounter-l2m1}
\end{figure}

\begin{figure}[!htbp]
    \centering
    \includegraphics[width=0.325\textwidth]{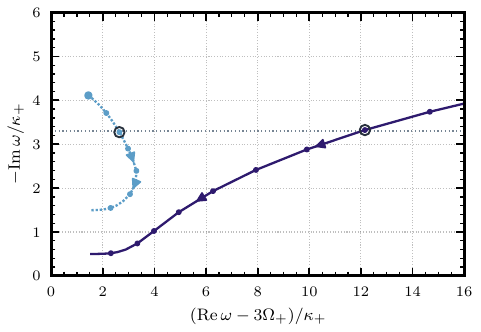}\hfill
    \includegraphics[width=0.325\textwidth]{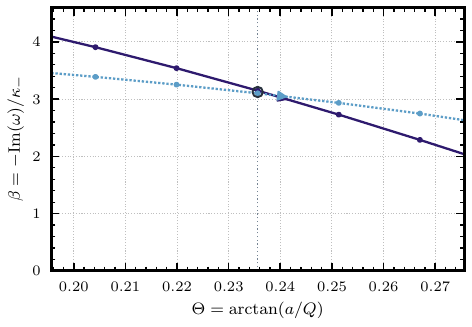}\hfill
    \includegraphics[width=0.325\textwidth]{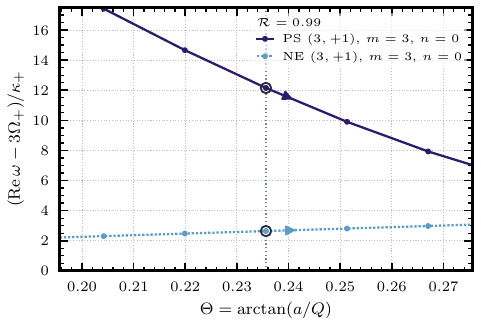}\\[0.25em]
    \includegraphics[width=0.325\textwidth]{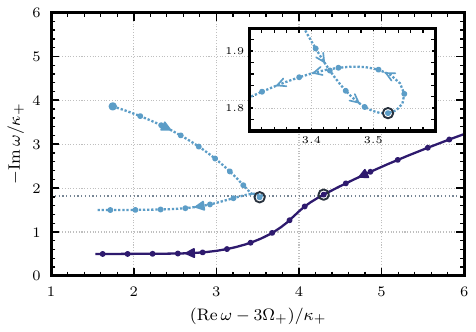}\hfill
    \includegraphics[width=0.325\textwidth]{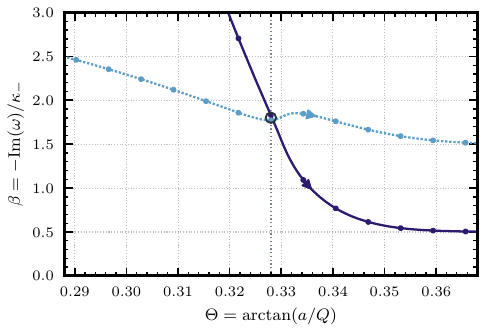}\hfill
    \includegraphics[width=0.325\textwidth]{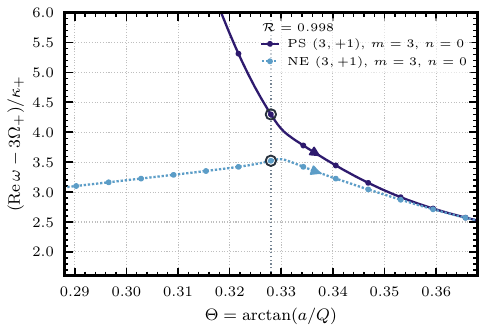}
    \caption{\textbf{The pair $(\ell,b)=(3,+1)$ with $m=3$}, as \cref{fig:intersection-repulsion}, for $\mathcal R=0.99$ (top) and $\mathcal R=0.998$ (bottom). At both values of $\mathcal R$ the damping rates cross, at $\Theta=0.239$ and $\Theta=0.328$, while the real parts differ: an intersection. Open circles mark the PS and NE modes at that value of $\Theta$ in all three panels, and they coincide in the middle panels, where the curves cross. Dotted lines mark the same value of $\Theta$. Beyond the range of $\Theta$ shown, the PS mode then settles on $N=0$ and the NE mode on $N=1$. At $\mathcal R=0.998$ the two modes pass much closer, and the NE mode makes a small loop in the complex plane, magnified in the inset.}
    \label{fig:encounter-l3m3}
\end{figure}

At $\mathcal R=0.998$ the two pairs of $(2,+1)$ behave differently. Their damping rates no longer cross and only their real parts do, which is an eigenvalue repulsion in the language of Ref.~\cite{Davey:2024xvd}. The near-extremal mode now stays the less damped of the two and ends on $N=0$, while the photon sphere mode ends two rungs higher, on $N=2$. \Cref{fig:intersection-repulsion} shows this for $m=2$, and \cref{fig:encounter-l2m1} shows the same behaviour for $m=1$. In both cases the rung in between, $N=1$, is taken by the near-extremal overtone with $n=1$, which we have also tracked and which stays between the other two modes in damping over the whole range of $\Theta$. The pair $(3,+1)$ with $m=3$ is still an intersection at $\mathcal R=0.998$, as \cref{fig:encounter-l3m3} shows, but the two modes now pass much closer to each other. At the value of $\Theta$ where their damping rates cross, their real parts differ by $9.07\,\kappa_+$ for $\mathcal R=0.99$ but only by $0.76\,\kappa_+$ for $\mathcal R=0.998$, and in the second case the near-extremal mode traces a small loop in the complex plane as the photon sphere mode goes past. Such loops also appear for the two pairs of $(2,+1)$ at $\mathcal R=0.99$, in the intersection stage just before these pairs turn into repulsions. For $m=2$ the loop is tiny, about $0.05\,\kappa_+$ across, and only shows up once the turn of the near-extremal mode is followed in steps of $\Delta\Theta=\pi/20000$. At $\mathcal R=0.998$, where the same two pairs repel, the near-extremal mode turns back without crossing its own path. The loops are seen in the coordinates we use for plotting, which are shifted by $m\Omega_+$ and measured in units of $\kappa_+$. The two modes never actually coincide. They come no closer than $0.57\,\kappa_+$, so there is no exceptional point on the real $\Theta$ axis.

\section{Convergence of the Spectral Method and Validation}\label{sm_sec:convergence}

\subsection{Convergence of the spectral method}

\begin{figure}[!htbp]
    \centering
    \includegraphics[width=\textwidth]{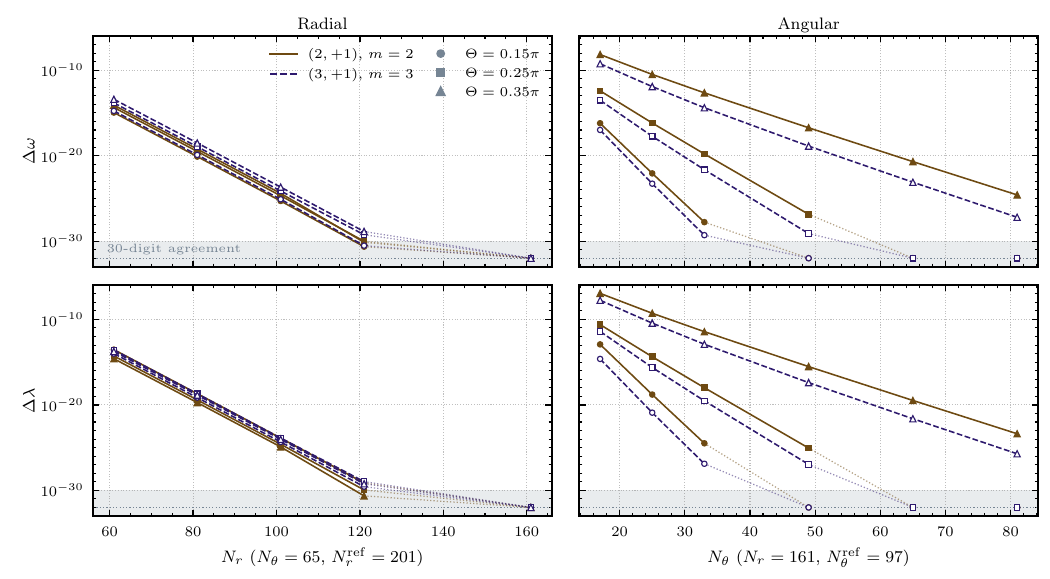}
    \caption{\textbf{Convergence of the spectral method.} Relative differences $\Delta\omega$ (top) and $\Delta\lambda$ (bottom) from the finest grid, for the photon sphere modes $(2,+1)$ with $m=2$ (solid lines, filled markers) and $(3,+1)$ with $m=3$ (dashed lines, open markers) of KNdS black holes with $y_+=1/2$ and $\mathcal R=0.95$, at $\Theta=0.15\pi$ (circles), $0.25\pi$ (squares), and $0.35\pi$ (triangles); the legend in the top-left panel is shared across all panels. Left: varying $N_r$ at fixed $N_\theta=65$; right: varying $N_\theta$ at fixed $N_r=161$. Eigenvalues are computed at 50-digit working precision with a Newton--Raphson tolerance of $10^{-28}$ and stored to 30 significant digits, showing exponential decay down to this storage limit. The shaded band marks agreement with the reference grid to all 30 stored digits, with faint dotted lines linking to the last nonzero points.}
    \label{fig:convergence}
\end{figure}

Throughout the main text, we monitor the convergence of the reported QNMs by comparing values across two $(N_r \times N_\theta)$ collocation grids: $81\times33$ and $101\times49$. To test the convergence rates of $\omega$ and $\lambda$ more systematically, here we vary the radial and angular resolutions independently. Focusing on the co-rotating photon sphere modes with $(\ell,b)=(2,+1)$, with $m=2$ and $(3,+1)$ with $m=3$ at $\Theta=0.15\pi$, $0.25\pi$, and $0.35\pi$, we recompute each mode across two grid sequences:
\begin{itemize}
    \item radial: $N_r = 61, 81, \dots, N^{\rm ref}_r = 201$ at fixed $N_\theta = 65$;
    \item angular: $N_\theta = 17, 25, \dots, N^{\rm ref}_\theta = 97$ at fixed $N_r = 161$.
\end{itemize}
On each grid, we follow the same refinement procedure as before: taking the value on the trajectory as an initial guess, we obtain seed eigenvectors from the full radial and angular eigenvalue problems and improve the root via Newton--Raphson iteration. Within each sequence, we compute the relative difference between the value on each grid and that on the finest reference grid ($N^{\rm ref}_r$ or $N^{\rm ref}_\theta$),
\begin{equation}
   \Delta \omega =  \left|1-\frac{\omega_N}{\omega^{\rm ref}}\right|,
\end{equation}
and analogously for $\lambda$, where $N$ denotes $N_r$ or $N_\theta$. As shown in \cref{fig:convergence}, the relative error falls linearly on a semi-logarithmic scale until it hits the round-off threshold set at $10^{-30}$. This confirms exponential convergence in both $N_r$ and $N_\theta$, a hallmark of pseudospectral methods.

\subsection{Limit to Reissner-Nordstr\"om-de~Sitter Black Hole}

In the limit $\Theta\to0$, the rotation parameter $a$ becomes zero and the BH reduces to a Reissner--Nordstr\"om--de Sitter (RNdS) BH with the same horizon radii $r_-$, $r_+$, and $r_{\rm c}$. The RNdS background is spherically symmetric, so its QNM frequencies do not depend on $m$. Moreover, each QNM trajectory of the KNdS BH should terminate at $\Theta=0$ on the RNdS mode whose label $(\ell,b,n)$ was assigned to it after solving the two-parameter eigenvalue problem at the beginning of our numerical recipe. Recall that the labels were assigned at $\Theta=5\pi/200$ by comparison with the RNdS spectrum, which we obtain from the decoupled master equations of Ref.~\cite{Stuchlik:2025mjj} with a standard spectral collocation method. The limit $\Theta\to0$ validates this assignment: each trajectory, when followed down to small $\Theta$, indeed ends on the mode to which it was assigned. We perform this validation step as follows.

Since the trajectories were computed only down to $\Theta=\pi/200$, we obtain their endpoints at $\Theta=0$ by extrapolation. For each trajectory, we take the polynomial of degree seven that interpolates its values at the eight points $\Theta=k\pi/200$, $k=1,\dots,8$, and evaluate it at $\Theta=0$. We estimate the error of this procedure by repeating it with only the first five, six, and seven points, and we take the largest resulting change in $\omega$ as the uncertainty. Since we solve for the counter-rotating modes with $m>0$, their trajectories have $\mathrm{Re}\,\omega<0$, so we compare them with $-\bar{\omega}$ of the corresponding RNdS mode. The results for all $27$ photon sphere trajectories are collected in \cref{tab:ps-rnds-limit}. Every trajectory ends on its assigned RNdS mode: the absolute difference between the extrapolated and RNdS frequencies is $\sim 10^{-8}$. Moreover, in each case the difference lies below the uncertainty of the extrapolation (which never exceeds $6\times10^{-7}$). We repeated the process for the near-extremal modes and found similar results.
\begin{table}[thbp]
\centering
\renewcommand{\arraystretch}{1.2}
\caption{The limit $\Theta\to0$ of the photon sphere trajectories of the main text, grouped by multiplet as in Fig.~\ref{fig:ps_modes_tracks_by_ell}, with $m$ as solved and the rotation as in Table~\ref{tab:gem-complete-spectrum}. $\omega r_{\rm c}$ at $\Theta\to0$: extrapolated from $\Theta=k\pi/200$, $k=1,\dots,8$, with its uncertainty, the largest change when 5 to 7 of these points are used. RNdS: the $n=0$ mode of the Reissner--Nordstr\"om--de Sitter multiplet to which the trajectory was assigned at $\Theta=5\pi/200$, from the master equations of Ref.~\cite{Stuchlik:2025mjj}, and the absolute difference from it; for a counter-rotating trajectory, whose $\mathrm{Re}\,\omega<0$, the difference is taken from $-\bar\omega$ of the RNdS mode. In every case the RNdS mode is also the nearest oscillatory RNdS mode with $\ell\ge|m|$.}
\label{tab:ps-rnds-limit}
\begin{ruledtabular}
\begin{tabular}{cccccc}
$m$ & rotation & $\omega r_{\rm c}$ at $\Theta\to0$ & uncertainty & RNdS $\omega r_{\rm c}$ & absolute difference \\
\hline
\multicolumn{6}{c}{\((\ell, b) = (1, -1)\), electromagnetic-led} \\
\hline
$1$ & counter & $-0.453609-0.068613i$ & $4\times10^{-7}$ & $0.453609-0.068613i$ & $1\times10^{-8}$ \\
$0$ & axisym. & $0.453609-0.068613i$ & $3\times10^{-7}$ & $0.453609-0.068613i$ & $9\times10^{-9}$ \\
$1$ & co & $0.453609-0.068613i$ & $2\times10^{-7}$ & $0.453609-0.068613i$ & $1\times10^{-9}$ \\
\hline
\multicolumn{6}{c}{\((\ell, b) = (2, +1)\), gravitational-led} \\
\hline
$2$ & counter & $-0.413878-0.072984i$ & $2\times10^{-7}$ & $0.413878-0.072984i$ & $2\times10^{-9}$ \\
$1$ & counter & $-0.413878-0.072984i$ & $4\times10^{-7}$ & $0.413878-0.072984i$ & $1\times10^{-8}$ \\
$0$ & axisym. & $0.413878-0.072984i$ & $3\times10^{-7}$ & $0.413878-0.072984i$ & $9\times10^{-9}$ \\
$1$ & co & $0.413878-0.072984i$ & $1\times10^{-7}$ & $0.413878-0.072984i$ & $1\times10^{-9}$ \\
$2$ & co & $0.413878-0.072984i$ & $5\times10^{-7}$ & $0.413878-0.072984i$ & $6\times10^{-8}$ \\
\hline
\multicolumn{6}{c}{\((\ell, b) = (2, -1)\), electromagnetic-led} \\
\hline
$2$ & counter & $-0.728010-0.070207i$ & $3\times10^{-7}$ & $0.728010-0.070207i$ & $3\times10^{-8}$ \\
$1$ & counter & $-0.728010-0.070207i$ & $2\times10^{-7}$ & $0.728010-0.070207i$ & $3\times10^{-8}$ \\
$0$ & axisym. & $0.728010-0.070207i$ & $3\times10^{-7}$ & $0.728010-0.070207i$ & $1\times10^{-8}$ \\
$1$ & co & $0.728010-0.070207i$ & $3\times10^{-7}$ & $0.728010-0.070207i$ & $8\times10^{-9}$ \\
$2$ & co & $0.728010-0.070207i$ & $2\times10^{-8}$ & $0.728010-0.070207i$ & $3\times10^{-9}$ \\
\hline
\multicolumn{6}{c}{\((\ell, b) = (3, +1)\), gravitational-led} \\
\hline
$3$ & counter & $-0.688797-0.072974i$ & $1\times10^{-7}$ & $0.688797-0.072974i$ & $7\times10^{-10}$ \\
$2$ & counter & $-0.688797-0.072974i$ & $3\times10^{-7}$ & $0.688797-0.072974i$ & $3\times10^{-8}$ \\
$1$ & counter & $-0.688797-0.072974i$ & $2\times10^{-7}$ & $0.688797-0.072974i$ & $3\times10^{-8}$ \\
$0$ & axisym. & $0.688797-0.072974i$ & $3\times10^{-7}$ & $0.688797-0.072974i$ & $1\times10^{-8}$ \\
$1$ & co & $0.688797-0.072974i$ & $3\times10^{-7}$ & $0.688797-0.072974i$ & $8\times10^{-9}$ \\
$2$ & co & $0.688797-0.072974i$ & $5\times10^{-8}$ & $0.688797-0.072974i$ & $3\times10^{-9}$ \\
$3$ & co & $0.688797-0.072974i$ & $2\times10^{-7}$ & $0.688797-0.072974i$ & $2\times10^{-8}$ \\
\hline
\multicolumn{6}{c}{\((\ell, b) = (3, -1)\), electromagnetic-led} \\
\hline
$3$ & counter & $-0.989603-0.070845i$ & $2\times10^{-7}$ & $0.989603-0.070845i$ & $3\times10^{-8}$ \\
$2$ & counter & $-0.989603-0.070845i$ & $5\times10^{-7}$ & $0.989603-0.070845i$ & $2\times10^{-8}$ \\
$1$ & counter & $-0.989603-0.070845i$ & $2\times10^{-7}$ & $0.989603-0.070845i$ & $2\times10^{-8}$ \\
$0$ & axisym. & $0.989603-0.070845i$ & $3\times10^{-7}$ & $0.989603-0.070845i$ & $1\times10^{-8}$ \\
$1$ & co & $0.989603-0.070845i$ & $4\times10^{-7}$ & $0.989603-0.070845i$ & $1\times10^{-8}$ \\
$2$ & co & $0.989603-0.070845i$ & $2\times10^{-7}$ & $0.989603-0.070845i$ & $9\times10^{-10}$ \\
$3$ & co & $0.989603-0.070845i$ & $1\times10^{-7}$ & $0.989603-0.070845i$ & $2\times10^{-9}$ \\
\end{tabular}
\end{ruledtabular}
\end{table}

\subsection{Comparison between direct integration and spectral methods}

\begin{figure}[!htbp]
    \centering
    \includegraphics[width=\textwidth]{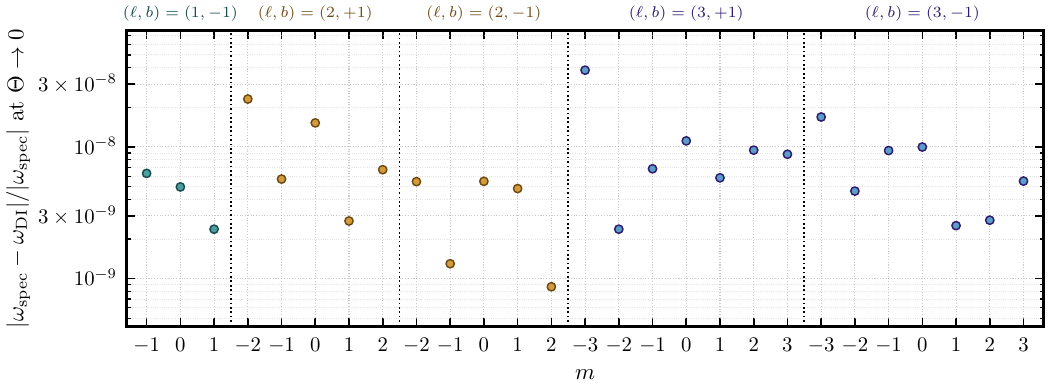}
    \caption{\textbf{Spectral method against direct integration at $\Theta\to0$} for the photon sphere trajectories of \cref{tab:ps-rnds-limit}, against $m$ (negative for counter-rotating modes).}
    \label{fig:ps-rnds-limit-absdiff}
\end{figure}

\begin{figure}[!htbp]
    \centering
    \includegraphics[width=\textwidth]{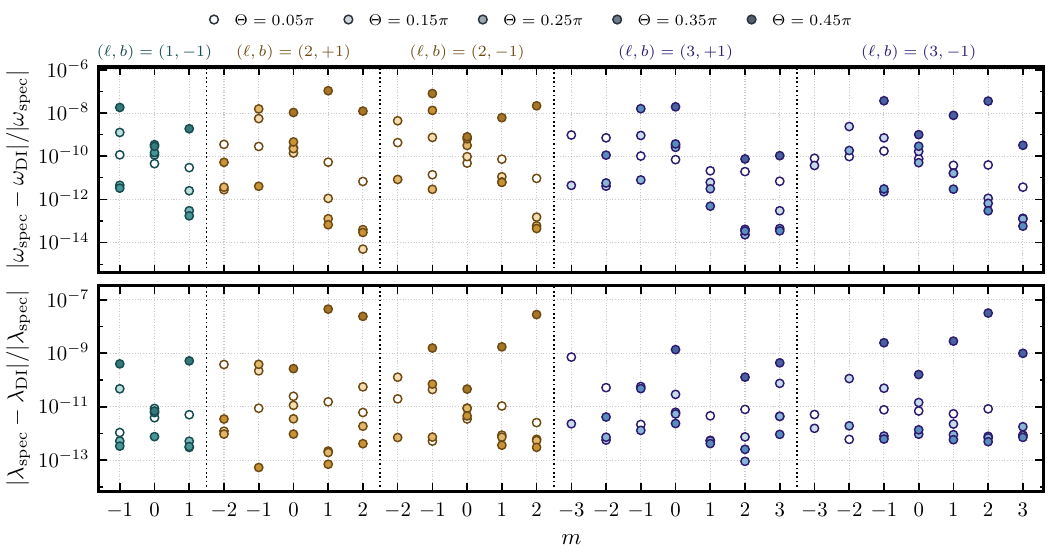}\par
    {\small (a) $n=0$ modes}\par
    \vspace{0.5cm}
    \includegraphics[width=\textwidth]{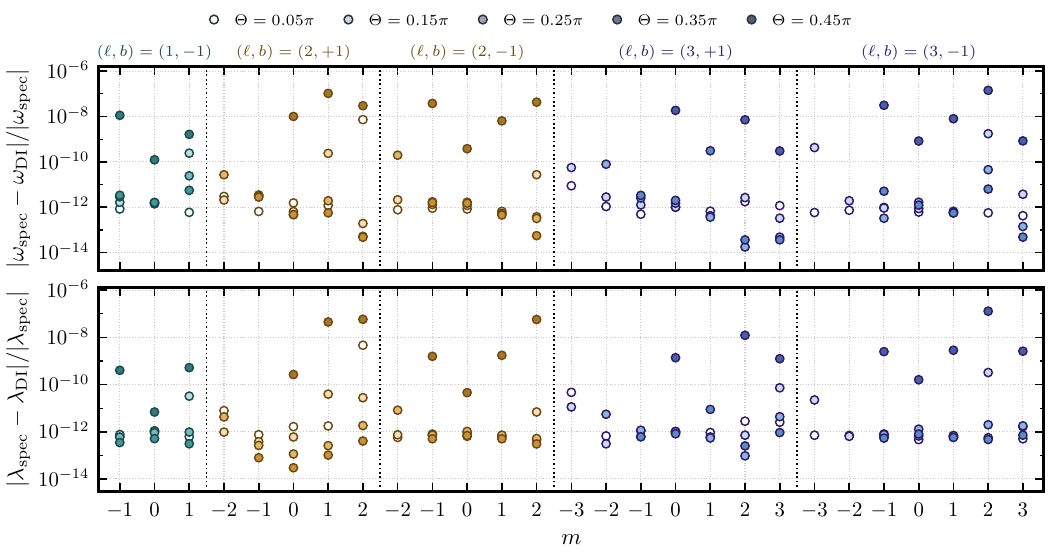}\par
    {\small (b) $n=1$ modes}
    \caption{\textbf{Spectral method against direct integration for the photon sphere modes} of \cref{tab:gem-complete-spectrum}: relative differences of $\omega$ and $\lambda$ against $m$ (negative for counter-rotating modes). Panel (a) shows the relative errors for the fundamental modes and panel (b) those for the first overtones.}
    \label{fig:comparison_ps_modes}
\end{figure}

\begin{figure}[!htbp]
    \centering
    \includegraphics[width=\textwidth]{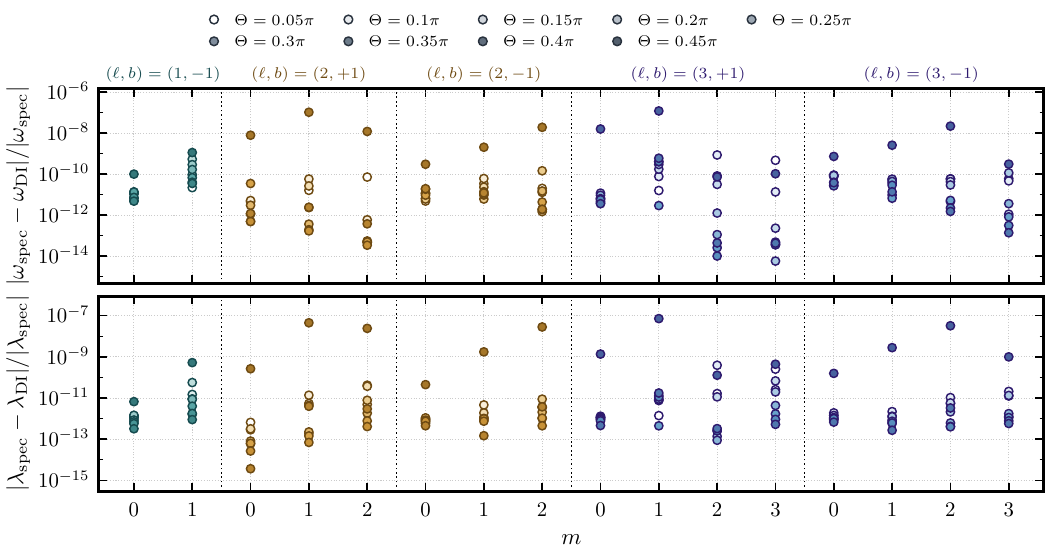}
    \caption{\textbf{Spectral method against direct integration for the near-extremal modes} of \cref{tab:zdm-spectrum}: relative differences of $\omega$  and $\lambda$.}
    \label{fig:comparison_zdm_modes}
\end{figure}

We compare the spectral method with the direct integration of Appendix~\ref{sec:directint} in two settings. For $a=0.2M$, $Q=0.4M$ and $\ell=m=2$, \cref{tab:gravrec,tab:emrec} list the frequencies and angular eigenvalues of the gravitational-led and electromagnetic-led modes obtained with both methods for several values of the cosmological constant $\Lambda=3/\Lds^2$, where $\Lds$ is the de~Sitter radius; the two methods agree to $\sim10^{-9}$. For the black holes with $y_+=1/2$ and $\mathcal R=0.95$, \cref{fig:ps-rnds-limit-absdiff} compares the values extrapolated to $\Theta\to0$ in \cref{tab:ps-rnds-limit} with direct integration, and \cref{fig:comparison_ps_modes,fig:comparison_zdm_modes} show the relative differences between the two methods for the photon sphere and near-extremal modes tabulated in \cref{tab:gem-complete-spectrum,tab:zdm-spectrum}. Across all converged modes shown in \cref{fig:ps-rnds-limit-absdiff,fig:comparison_ps_modes,fig:comparison_zdm_modes}, the relative differences between the two methods remain $\lesssim 10^{-7}$, well below the six-decimal rounding level of \cref{tab:ps-rnds-limit,tab:gem-complete-spectrum,tab:zdm-spectrum}.


\begin{table}[thbp]
\centering
\renewcommand{\arraystretch}{1.2}
\setlength{\tabcolsep}{3pt}
\footnotesize
\caption{Spectral method against direct integration for the gravitational-led modes ($b=+1$) with $a=0.2M$, $Q=0.4M$ and $\ell=m=2$, at several values of the de~Sitter radius $\Lds$, $\Lambda=3/\Lds^2$. For the quasinormal frequency $M\omega$ and the angular eigenvalue $\lambda$, we list the values from both methods, their absolute difference, and the relative differences $\delta_{\rm Re}$ and $\delta_{\rm Im}$ of their real and imaginary parts.}
\label{tab:gravrec}
\begin{tabular}{@{}w{c}{0.07\dimexpr\textwidth-30pt\relax}w{r}{0.125\dimexpr\textwidth-30pt\relax}@{}w{l}{0.14\dimexpr\textwidth-30pt\relax}w{r}{0.125\dimexpr\textwidth-30pt\relax}@{}w{l}{0.14\dimexpr\textwidth-30pt\relax}w{c}{0.17\dimexpr\textwidth-30pt\relax}w{c}{0.115\dimexpr\textwidth-30pt\relax}w{c}{0.115\dimexpr\textwidth-30pt\relax}@{}}
\hline\hline
\(\Lds/M\) & \multicolumn{2}{c}{Spectral method} & \multicolumn{2}{c}{Direct integration} & Absolute difference & \(\delta_{\rm Re}\) & \(\delta_{\rm Im}\) \\
\hline
\multicolumn{8}{c}{Quasinormal frequency \(M\omega\)} \\
\hline
$15$ & $0.389936072$ & ${}-0.084227083i$ & $0.389936072$ & ${}-0.084227083i$ & $4.4\times10^{-10}$ & $1.1\times10^{-9}$ & $6.7\times10^{-10}$ \\
$20$ & $0.398840564$ & ${}-0.086193397i$ & $0.398840564$ & ${}-0.086193397i$ & $5.1\times10^{-10}$ & $4.9\times10^{-10}$ & $5.5\times10^{-9}$ \\
$30$ & $0.405044147$ & ${}-0.087552162i$ & $0.405044147$ & ${}-0.087552162i$ & $3.5\times10^{-10}$ & $8.6\times10^{-10}$ & $5.0\times10^{-10}$ \\
$40$ & $0.407186052$ & ${}-0.088019161i$ & $0.407186052$ & ${}-0.088019161i$ & $3.6\times10^{-10}$ & $8.6\times10^{-10}$ & $1.0\times10^{-9}$ \\
$60$ & $0.408706927$ & ${}-0.088350087i$ & $0.408706927$ & ${}-0.088350087i$ & $4.1\times10^{-10}$ & $8.4\times10^{-10}$ & $2.6\times10^{-9}$ \\
$80$ & $0.409237471$ & ${}-0.088465397i$ & $0.409237471$ & ${}-0.088465397i$ & $1.6\times10^{-10}$ & $3.0\times10^{-10}$ & $1.1\times10^{-9}$ \\
\hline
\multicolumn{8}{c}{Angular eigenvalue \(\lambda\)} \\
\hline
$15$ & $4.171477860$ & ${}+0.017000771i$ & $4.171477860$ & ${}+0.017000771i$ & $4.5\times10^{-10}$ & $7.7\times10^{-11}$ & $1.9\times10^{-8}$ \\
$20$ & $4.169550659$ & ${}+0.017395288i$ & $4.169550659$ & ${}+0.017395288i$ & $6.5\times10^{-10}$ & $1.2\times10^{-10}$ & $2.5\times10^{-8}$ \\
$30$ & $4.168205950$ & ${}+0.017667810i$ & $4.168205950$ & ${}+0.017667810i$ & $3.1\times10^{-10}$ & $5.7\times10^{-11}$ & $1.1\times10^{-8}$ \\
$40$ & $4.167741267$ & ${}+0.017761456i$ & $4.167741267$ & ${}+0.017761456i$ & $3.2\times10^{-10}$ & $3.3\times10^{-11}$ & $1.6\times10^{-8}$ \\
$60$ & $4.167411192$ & ${}+0.017827809i$ & $4.167411192$ & ${}+0.017827809i$ & $1.7\times10^{-10}$ & $1.1\times10^{-11}$ & $9.1\times10^{-9}$ \\
$80$ & $4.167296024$ & ${}+0.017850929i$ & $4.167296024$ & ${}+0.017850929i$ & $5.0\times10^{-10}$ & $5.5\times10^{-11}$ & $2.5\times10^{-8}$ \\
\hline\hline
\end{tabular}
\end{table}

\begin{table}[thbp]
\centering
\renewcommand{\arraystretch}{1.2}
\setlength{\tabcolsep}{3pt}
\footnotesize
\caption{Same as Table~\ref{tab:gravrec}, for the electromagnetic-led modes ($b=-1$). The direct-integration values are those for $\varsigma=-1$; the values for $\varsigma=+1$ agree with them to better than $10^{-9}$.}
\label{tab:emrec}
\begin{tabular}{@{}w{c}{0.07\dimexpr\textwidth-30pt\relax}w{r}{0.125\dimexpr\textwidth-30pt\relax}@{}w{l}{0.14\dimexpr\textwidth-30pt\relax}w{r}{0.125\dimexpr\textwidth-30pt\relax}@{}w{l}{0.14\dimexpr\textwidth-30pt\relax}w{c}{0.17\dimexpr\textwidth-30pt\relax}w{c}{0.115\dimexpr\textwidth-30pt\relax}w{c}{0.115\dimexpr\textwidth-30pt\relax}@{}}
\hline\hline
\(\Lds/M\) & \multicolumn{2}{c}{Spectral method} & \multicolumn{2}{c}{Direct integration} & Absolute difference & \(\delta_{\rm Re}\) & \(\delta_{\rm Im}\) \\
\hline
\multicolumn{8}{c}{Quasinormal frequency \(M\omega\)} \\
\hline
$15$ & $0.489254738$ & ${}-0.090072251i$ & $0.489254736$ & ${}-0.090072251i$ & $1.6\times10^{-9}$ & $3.3\times10^{-9}$ & $1.3\times10^{-9}$ \\
$20$ & $0.500748600$ & ${}-0.092517838i$ & $0.500748599$ & ${}-0.092517838i$ & $1.4\times10^{-9}$ & $2.8\times10^{-9}$ & $2.9\times10^{-9}$ \\
$30$ & $0.508756495$ & ${}-0.094226292i$ & $0.508756494$ & ${}-0.094226292i$ & $1.4\times10^{-9}$ & $2.8\times10^{-9}$ & $8.5\times10^{-10}$ \\
$40$ & $0.511521478$ & ${}-0.094817087i$ & $0.511521477$ & ${}-0.094817087i$ & $1.5\times10^{-9}$ & $2.9\times10^{-9}$ & $4.6\times10^{-9}$ \\
$60$ & $0.513484815$ & ${}-0.095236878i$ & $0.513484813$ & ${}-0.095236878i$ & $1.6\times10^{-9}$ & $3.1\times10^{-9}$ & $4.4\times10^{-9}$ \\
$80$ & $0.514169717$ & ${}-0.095383376i$ & $0.514169716$ & ${}-0.095383376i$ & $1.4\times10^{-9}$ & $2.6\times10^{-9}$ & $3.2\times10^{-9}$ \\
\hline
\multicolumn{8}{c}{Angular eigenvalue \(\lambda\)} \\
\hline
$15$ & $-4.183277042$ & ${}-0.012089645i$ & $-4.183277038$ & ${}-0.012089645i$ & $4.1\times10^{-9}$ & $9.7\times10^{-10}$ & $8.8\times10^{-9}$ \\
$20$ & $-4.181672640$ & ${}-0.012406146i$ & $-4.181672635$ & ${}-0.012406146i$ & $4.7\times10^{-9}$ & $1.1\times10^{-9}$ & $3.1\times10^{-8}$ \\
$30$ & $-4.180554696$ & ${}-0.012626878i$ & $-4.180554692$ & ${}-0.012626878i$ & $4.3\times10^{-9}$ & $1.0\times10^{-9}$ & $2.4\times10^{-8}$ \\
$40$ & $-4.180168663$ & ${}-0.012703137i$ & $-4.180168659$ & ${}-0.012703137i$ & $3.9\times10^{-9}$ & $9.2\times10^{-10}$ & $2.8\times10^{-9}$ \\
$60$ & $-4.179894544$ & ${}-0.012757301i$ & $-4.179894540$ & ${}-0.012757301i$ & $4.1\times10^{-9}$ & $9.8\times10^{-10}$ & $6.0\times10^{-9}$ \\
$80$ & $-4.179798917$ & ${}-0.012776199i$ & $-4.179798913$ & ${}-0.012776199i$ & $4.2\times10^{-9}$ & $9.9\times10^{-10}$ & $3.0\times10^{-8}$ \\
\hline\hline
\end{tabular}
\end{table}

\section{The de Sitter Modes}
\label{app:dS_modes}
\begin{figure}[!htbp]
    \centering
    \includegraphics[width=\textwidth]{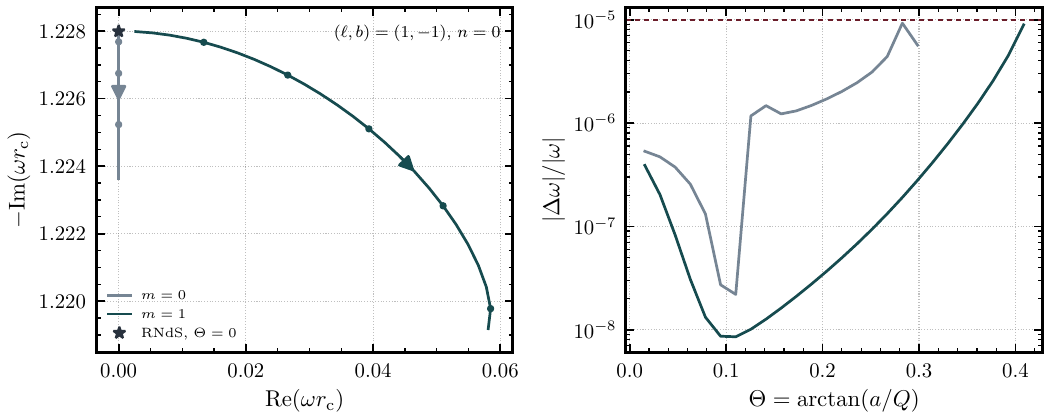}
    \caption{\textbf{The least damped de Sitter modes}, $(\ell,b)=(1,-1)$ with $m=0$ and $1$, for $y_+=1/2$ and $\mathcal R=0.95$. \textbf{Left:} trajectories from the RNdS mode at $\Theta=0$ (star), with arrowheads pointing towards increasing $\Theta$, shown while the two collocation grids agree. \textbf{Right:} the relative difference of $\omega$ between the $81\times33$ and $101\times49$ grids.}
    \label{fig:ds-l1-family}
\end{figure}

Besides the photon sphere and near-extremal families, the spectrum contains a third family, the de Sitter modes, which are tied to the cosmological horizon
~\cite{Stuchlik:2025mjj,Cardoso:2017soq}.
For the coupled perturbations, the lowest-lying de Sitter family is therefore the single multiplet $(\ell,b)=(1,-1)$ with $n=0$, whose three members $m=-1,0,1$ are degenerate at $\Theta=0$, with a purely imaginary frequency.  At a working precision of 64 digits, we were able to track the lowest-lying de Sitter mode. In \cref{fig:ds-l1-family} we show the two trajectories up to the values of $\Theta$ at which they had converged, that is, at which the computations on $81\times33$ and $101\times49$ points still agreed to $10^{-5}$: $\Theta=0.095\pi$ for $m=0$ and $\Theta=0.13\pi$ for $m=1$. Beyond these values, we could not see convergence. Although they do not play any role in SCC, they may display interesting dynamics in the complex plane, potentially showing bifurcation and eigenvalue repulsion. We wish to investigate this family in detail in a forthcoming work. 

\FloatBarrier
\section{Numerical Values of Gravito-electromagnetic Quasinormal Modes} \label{app:numerical_values_gem_qnms} 
 We tabulate selected gravito-electromagnetic QNM frequencies $\omega r_{\rm c}$ of the Kerr--Newman--de~Sitter black hole, computed with the spectral method used in the main text, for $\Theta = k \pi/200$, with $k = 10,30,50,70,90$. The values for the photon sphere family of modes are given in \cref{tab:gem-complete-spectrum}. The table includes the fundamental modes as well as the first overtones for the different Reissner--Nordstr\"om--de Sitter
multiplets $(\ell,b)$. For each value of $\Theta$, the members from $m=-\ell$ to $\ell$ are given.
Similarly, \cref{tab:zdm-spectrum} lists the values of the near-extremal QNMs at $\Theta=k\pi/200$, $k=10,20,\dots,90$. Here each multiplet $(\ell,b)$ has only the members $m=0,\dots,\ell$. 

\begingroup
\setlength{\LTcapwidth}{\textwidth}
\renewcommand{\arraystretch}{1.2}
\setlength{\tabcolsep}{3pt}
\footnotesize
\begin{longtable}{@{}w{c}{4.1em}w{c}{4.1em}w{c}{4.1em}w{c}{4.1em}w{c}{4.1em}c@{}}
\caption{Photon sphere gravito-electromagnetic quasinormal modes, the fundamental modes ($n=0$) and the first overtones ($n=1$), of Kerr--Newman--de Sitter black holes with $y_+=r_+/r_{\rm c}=1/2$ and $\mathcal R=0.95$, $\mathcal R^2=r_-/r_+$, for the spin label $\varsigma=+1$, in units $r_{\rm c}=1$: the frequencies $\omega r_{\rm c}$ and separation constants $\lambda$ at $\Theta=k\pi/200$, $k=10,30,50,70,90$, every second point marked on the curves of Figs.~\ref{fig:ps_modes_tracks_by_ell} and~\ref{fig:ps_n0_n1_overtone_tracks_by_ell}. The background parameters $a$, $Q$, $M$ and $\Lambda$ at each $\Theta$ follow from Eq.~\eqref{eq:MaQ}. The table has one section per Reissner--Nordstr\"om--de Sitter multiplet $(\ell,b)$ from which the modes continue, with $b=+1$ the gravitational-led and $b=-1$ the electromagnetic-led branch, in the order of the panels of Fig.~\ref{fig:ps_modes_tracks_by_ell}. At each $\Theta$ the members of a multiplet are listed as the curves of its panel, from $m=-\ell$ to $\ell$: first the counter-rotating modes with $m=\ell,\dots,1$ ($\mathrm{Re}\,\omega<0$ at small $\Theta$; they are the $m<0$ modes at $-\bar\omega$), then the axisymmetric mode $m=0$, then the co-rotating modes with $m=1,\dots,\ell$; the fundamental modes come first and, below a rule, the first overtones of the same members, for the same black hole. The values of $\omega$ and $\lambda$ are those of the equations solved with $m\ge0$. A dagger marks the frequencies at which the two collocation grids differ by more than $10^{-5}$ in $\omega$ (the faint parts of the curves); there the value only continues the track.\vspace{2ex}}
\label{tab:gem-complete-spectrum} \\
\hline\hline
\(\Theta\) & \(a\) & \(Q\) & \(M\) & \(\Lambda\) & \begin{tabular}{@{}w{c}{1.4em} w{c}{3.6em} w{c}{2.8em} w{r}{4.6em}@{}w{l}{6.0em} w{r}{4.6em}@{}w{l}{5.6em}@{}}\(m\) & rotation & \(n\) & \multicolumn{2}{w{c}{10.6em}}{\(\omega r_{\rm c}\)} & \multicolumn{2}{w{c}{10.2em}@{}}{\(\lambda\)}\end{tabular} \\
\hline
\endfirsthead
\caption[]{Photon sphere gravito-electromagnetic quasinormal modes (continued).\vspace{2ex}} \\
\hline\hline
\(\Theta\) & \(a\) & \(Q\) & \(M\) & \(\Lambda\) & \begin{tabular}{@{}w{c}{1.4em} w{c}{3.6em} w{c}{2.8em} w{r}{4.6em}@{}w{l}{6.0em} w{r}{4.6em}@{}w{l}{5.6em}@{}}\(m\) & rotation & \(n\) & \multicolumn{2}{w{c}{10.6em}}{\(\omega r_{\rm c}\)} & \multicolumn{2}{w{c}{10.2em}@{}}{\(\lambda\)}\end{tabular} \\
\hline
\endhead
\hline
\endfoot
\hline\hline
\endlastfoot
\multicolumn{6}{@{}c@{}}{\begin{tabular}{@{}w{c}{4.1em}w{c}{4.1em}w{c}{4.1em}w{c}{4.1em}w{c}{4.1em}c@{}}
\multicolumn{6}{c}{\((\ell, b) = (1, -1)\), electromagnetic-led} \\
\hline
0.157080 & 0.063948 & 0.403751 & 0.392993 & 1.138697 & \begin{tabular}[c]{@{}w{c}{1.4em} w{c}{3.6em} w{c}{2.8em} w{r}{4.6em}@{}w{l}{6.0em} w{r}{4.6em}@{}w{l}{5.6em}@{}}
$1$ & counter &  & $-0.328929$ & ${}-0.073778i$ & $2.228402$ & ${}-0.004520i$ \\
$0$ & axisym. & $n=0$ & $0.453365$ & ${}-0.068047i$ & $2.581352$ & ${}-0.026489i$ \\
$1$ & co &  & $0.587034$ & ${}-0.060371i$ & $2.664924$ & ${}-0.048559i$ \\
\hline
$1$ & counter &  & $-0.321614$ & ${}-0.222344i$ & $2.205999$ & ${}-0.015512i$ \\
$0$ & axisym. & $n=1$ & $0.443081$ & ${}-0.205666i$ & $2.562482$ & ${}-0.078970i$ \\
$1$ & co &  & $0.572591$ & ${}-0.183026i$ & $2.645446$ & ${}-0.146368i$ \\
\end{tabular}
\end{tabular}} \\
\hline
0.471239 & 0.184537 & 0.362175 & 0.388574 & 1.125891 & \begin{tabular}[c]{@{}w{c}{1.4em} w{c}{3.6em} w{c}{2.8em} w{r}{4.6em}@{}w{l}{6.0em} w{r}{4.6em}@{}w{l}{5.6em}@{}}
$1$ & counter &  & $-0.129625$ & ${}-0.075145i$ & $2.499605$ & ${}+0.046612i$ \\
$0$ & axisym. & $n=0$ & $0.440504$ & ${}-0.063951i$ & $2.680093$ & ${}-0.026830i$ \\
$1$ & co &  & $0.822560$ & ${}-0.040488i$ & $2.376885$ & ${}-0.015818i$ \\
\hline
$1$ & counter &  & $-0.126178$ & ${}-0.225814i$ & $2.478237$ & ${}+0.140764i$ \\
$0$ & axisym. & $n=1$ & $0.431399$ & ${}-0.193268i$ & $2.661699$ & ${}-0.080081i$ \\
$1$ & co &  & $0.803098$ & ${}-0.121099i$ & $2.362725$ & ${}-0.045050i$ \\
\end{tabular} \\
\end{longtable}
\endgroup

\begingroup
\setlength{\LTcapwidth}{\textwidth}
\renewcommand{\arraystretch}{1.2}
\setlength{\tabcolsep}{3pt}
\footnotesize
\begin{longtable}{@{}w{c}{4.1em}w{c}{4.1em}w{c}{4.1em}w{c}{4.1em}w{c}{4.1em}c@{}}
\caption{Near-extremal gravito-electromagnetic quasinormal modes of Kerr--Newman--de Sitter black holes with $y_+=r_+/r_{\rm c}=1/2$ and $\mathcal R=0.95$, $\mathcal R^2=r_-/r_+$, for the spin label $\varsigma=+1$, in units $r_{\rm c}=1$: the frequencies $\omega r_{\rm c}$ and separation constants $\lambda$ at $\Theta=k\pi/200$, $k=10,20,\dots,90$, the points marked on the curves of the top row of Fig.~\ref{fig:zdm-tracks-all}. The background parameters $a$, $Q$, $M$ and $\Lambda$ at each $\Theta$ follow from Eq.~\eqref{eq:MaQ}. The table has one section per Reissner--Nordstr\"om--de Sitter multiplet $(\ell,b)$ on the zero-damped branch from which the modes continue, and at each $\Theta$ the members are listed by $m$; a near-extremal multiplet has the members $m=0,\dots,\ell$ only. For $m=0$ the frequency is purely imaginary.\vspace{2ex}}
\label{tab:zdm-spectrum} \\
\hline\hline
\(\Theta\) & \(a\) & \(Q\) & \(M\) & \(\Lambda\) & \begin{tabular}{@{}w{c}{1.4em} w{c}{3.6em} w{c}{2.8em} w{r}{4.6em}@{}w{l}{6.0em} w{r}{4.6em}@{}w{l}{5.6em}@{}}\(m\) & rotation & \(n\) & \multicolumn{2}{w{c}{10.6em}}{\(\omega r_{\rm c}\)} & \multicolumn{2}{w{c}{10.2em}@{}}{\(\lambda\)}\end{tabular} \\
\hline
\endfirsthead
\caption[]{Near-extremal gravito-electromagnetic quasinormal modes (continued).\vspace{2ex}} \\
\hline\hline
\(\Theta\) & \(a\) & \(Q\) & \(M\) & \(\Lambda\) & \begin{tabular}{@{}w{c}{1.4em} w{c}{3.6em} w{c}{2.8em} w{r}{4.6em}@{}w{l}{6.0em} w{r}{4.6em}@{}w{l}{5.6em}@{}}\(m\) & rotation & \(n\) & \multicolumn{2}{w{c}{10.6em}}{\(\omega r_{\rm c}\)} & \multicolumn{2}{w{c}{10.2em}@{}}{\(\lambda\)}\end{tabular} \\
\hline
\endhead
\hline
\endfoot
\hline\hline
\endlastfoot
\multicolumn{6}{@{}c@{}}{\begin{tabular}{@{}w{c}{4.1em}w{c}{4.1em}w{c}{4.1em}w{c}{4.1em}w{c}{4.1em}c@{}}
\multicolumn{6}{c}{\((\ell, b) = (1, -1)\), electromagnetic-led} \\
\hline
0.157080 & 0.063948 & 0.403751 & 0.392993 & 1.138697 & \begin{tabular}[c]{@{}w{c}{1.4em} w{c}{3.6em} w{c}{2.8em} w{r}{4.6em}@{}w{l}{6.0em} w{r}{4.6em}@{}w{l}{5.6em}@{}}
$0$ & axisym. &  & $0.000000$ & ${}-0.193996i$ & $2.476076$ & ${}+0.000000i$ \\
$1$ & co & $n=0$ & $0.283850$ & ${}-0.190371i$ & $2.435993$ & ${}-0.118153i$ \\
\end{tabular}
\end{tabular}} \\
\hline
0.314159 & 0.126039 & 0.387909 & 0.391241 & 1.133621 & \begin{tabular}[c]{@{}w{c}{1.4em} w{c}{3.6em} w{c}{2.8em} w{r}{4.6em}@{}w{l}{6.0em} w{r}{4.6em}@{}w{l}{5.6em}@{}}
$0$ & axisym. &  & $0.000000$ & ${}-0.183529i$ & $2.503588$ & ${}+0.000000i$ \\
$1$ & co & $n=0$ & $0.531661$ & ${}-0.171064i$ & $2.363563$ & ${}-0.101658i$ \\
\end{tabular} \\
\hline
0.471239 & 0.184537 & 0.362175 & 0.388574 & 1.125891 & \begin{tabular}[c]{@{}w{c}{1.4em} w{c}{3.6em} w{c}{2.8em} w{r}{4.6em}@{}w{l}{6.0em} w{r}{4.6em}@{}w{l}{5.6em}@{}}
$0$ & axisym. &  & $0.000000$ & ${}-0.169349i$ & $2.573200$ & ${}+0.000000i$ \\
$1$ & co & $n=0$ & $0.723093$ & ${}-0.146930i$ & $2.332359$ & ${}-0.042280i$ \\
\end{tabular} \\
\hline
0.628319 & 0.237917 & 0.327465 & 0.385313 & 1.116443 & \begin{tabular}[c]{@{}w{c}{1.4em} w{c}{3.6em} w{c}{2.8em} w{r}{4.6em}@{}w{l}{6.0em} w{r}{4.6em}@{}w{l}{5.6em}@{}}
$0$ & axisym. &  & $0.000000$ & ${}-0.154386i$ & $2.682777$ & ${}+0.000000i$ \\
$1$ & co & $n=0$ & $0.854087$ & ${}-0.123058i$ & $2.387924$ & ${}-0.001530i$ \\
\end{tabular} \\
\hline
0.785398 & 0.284915 & 0.284915 & 0.381821 & 1.106326 & \begin{tabular}[c]{@{}w{c}{1.4em} w{c}{3.6em} w{c}{2.8em} w{r}{4.6em}@{}w{l}{6.0em} w{r}{4.6em}@{}w{l}{5.6em}@{}}
$0$ & axisym. &  & $0.000000$ & ${}-0.140683i$ & $2.856477$ & ${}+0.000000i$ \\
$1$ & co & $n=0$ & $0.932657$ & ${}-0.105685i$ & $2.563192$ & ${}+0.015061i$ \\
\end{tabular} \\
\end{longtable}
\endgroup

\clearpage
\twocolumngrid
\bibliographystyle{custom_apsrev4-2_full_title}
\bibliography{references}

\end{document}